\documentclass[12pt,english,american]{article}

\usepackage{amsmath,amsfonts,amssymb,amsthm,mathrsfs}

\usepackage[colorlinks=true,linkcolor=Mahogany,citecolor=NavyBlue]{hyperref}

\usepackage{setspace}

\usepackage{enumerate}
\usepackage{bm}
\usepackage{bbm}
\usepackage{verbatim}
\usepackage[usenames,dvipsnames]{color}
\usepackage{empheq}

\theoremstyle{plain}\newtheorem{theorem}{Theorem}[section]
\theoremstyle{plain}\newtheorem{lemma}[theorem]{Lemma}
\theoremstyle{plain}
\theoremstyle{plain}
\theoremstyle{plain}\newtheorem{proposition}[theorem]{Proposition}
\theoremstyle{definition}
\theoremstyle{definition}
\theoremstyle{definition}\newtheorem{def:and:lemma}[theorem]{Definition and Lemma}

\newcommand{\id}{\mathbbm{1}}

\newcommand{\T}[1]{}

\usepackage[colorinlistoftodos,prependcaption,textsize=tiny]{todonotes}

\newcounter{remarks}
\newcommand{\lsp}{\big \langle }
\newcommand{\rsp}{\big \rangle }
\newcommand{\sno}{\small \vert\hspace{-1pt}\small \vert}
\newcommand{\D}{\textnormal{d}} 

\title{Effective pair interaction between impurity particles\\ induced by a dense Fermi gas}

\author{David Mitrouskas and Peter Pickl}

\date{}

\begin{document}

\maketitle

\frenchspacing

\begin{spacing}{1.15} 

\begin{abstract}
We study the dynamics of a small number of impurity particles coupled to the ideal Fermi gas in a $d$-dimensional box. The impurities interact with the fermions via a two-body potential $\lambda v(x)$ where $\lambda$ is a coupling constant and $v(x)$ for instance a screened Coulomb potential. After taking the large-volume limit at positive Fermi momentum $k_F$ we consider the regime of high density of the fermions, that is, $k_F$ large compared to one. For coupling constants that scale like $\lambda^2 \sim k_F^{(2-d)}$ we show that the impurity particles effectively decouple from the fermions but evolve with an attractive pair interaction among each other which is induced by fluctuations in the Fermi gas.
\end{abstract}

\section{Introduction and Main Result}

The presence of a surrounding medium can change the behaviour of quantum particles in a drastic way. Besides altering individual properties like mass and charge, the presence of an enviroment might also lead to medium-induced interactions between the particles. A prominent example of this effect is the phonon-mediated interaction between two repulsive polarons. The induced interaction is attractive and has the potential to overcompensate the repulsion between the polarons which may even cause the formation of a new quasi particle, the so-called bipolaron \cite{Camacho,Devreese2009,Franketal2012}. Other examples arise in the theory of ultra cold atoms, for instance, Casimir-type forces between heavy fermions placed into a Fermi sea \cite{Nishida2009} or effective interactions between Fermi polarons \cite{EnsEtAl2020,MistakidisEtAl18} or angulons \cite{SchmidtEtAl2020}. The effect of fermion-mediated interactions is known also for dilute mixtures of Bose-Fermi gases \cite{Huang20,Kinnunen15,Santamore08} for which experimental observations have been reported in \cite{DeSalvoEtAl1029,EdriEtAl2020}.

In this article we are interested in medium-induced interactions for a system of impurity particles immersed into a dense ideal Fermi gas. To this end we analyze the many-body Schr\"odinger time-evolution of $n\ge 2$ impurity particles coupled to a large number of fermions via a suitable two-body potential, for instance a screened Coulomb potential. Our main result shows that the impurity particles effectively decouple from the fermions if the number of fermions (per unit volume) becomes large. The presence of the Fermi gas, however, leaves its trace as an attractive interaction between the impurities. Physically, one can think of the induced interaction as being mediated by the creation and annihilation of electron-hole pairs in the Fermi sea.

This work generalizes our previous findings in \cite{JeblickMPP2017,JeblickMP2018}. While \cite{JeblickMPP2017} studies a single impurity that effectively decouples from the Fermi gas, in \cite{JeblickMP2018} we considered the case $n=2$ and derived the emergence of an effective interaction. Apart from generalzing to $n\ge 3$, our present analysis adds several other important improvements: (i) We extend the result to three spatial dimensions (in \cite{JeblickMPP2017,JeblickMP2018} we focused on the two-dimensional case). (ii) We treat pair potentials with a Coulomb singularity whereas our earlier results were restricted to bounded potentials. (iii) We obtain improved error estimates and a more transparent proof. (iv) Most importantly, in our opinion, we discuss the effective interaction among the impurities in more detail. This allows us to show that it adds a non-trivial effect at leading order to the effective dynamics (see Proposition \ref{prop: lower bound for h_0}).

The article is organized as follows. In the next two sections we introduce the microscopic and effective models, respectively. In Section \ref{sec: main results} we state our main results in which we compare the time-evolved states of the microscopic model and the effective model. All proofs are postponed to Section \ref{sec: proofs}.  

\subsection{The model}

We consider a system of $n \ge 2$ impurity particles and $N$ fermions in a $d$-dimensional cube $\Lambda  = [0,L]^d$ with periodic boundary conditions. To this model we assign the Hilbert space $\mathscr{H}_n \otimes   \mathscr H_N^- $ where $\mathscr{H}_n  = L^2(\Lambda )^{\otimes n}$ describes the states of the impurities with coordinates $y_1,\ldots, y_n  $ and $\mathscr H_N^- = \bigwedge^{N} L^2( \Lambda  )$ (the subspace of all anti-symmetric wave functions in $\mathscr H_N$) is the state space for the fermions with coordinates $x_1,\ldots x_N$. The Hamiltonian is given by
\begin{align}\label{eq: Hamiltonian}
H & =  \underbrace{\sum_{i =1}^n (- \Delta_{y_i } )+  \sum_{i < j}^n w(y_i - y_j )}_{= :\  h^0_n} + \underbrace{\sum_{i=1}^N ( -\Delta_{x_i})  + \lambda \sum_{i=1}^n \sum_{j=1}^N   v_L (  y_i - x _j)  }_{=:\ T_N \,+ \, V_{N+n}}  
\end{align}
with $h^0_n$ and $T_N$ acting only on the tensor components $\mathscr H_n$ and $\mathscr H_N^-$, respectively, that is, they have to be understood as $h^0_n \otimes \id $ and $\id \otimes T_N$. For the pair potentials $w$ and $v_L$ we suppose the following properties.
\begin{itemize}
\item[\hypertarget{(A$w$)}{(A$w$)}] $w$ is a real-valued even function on $\Lambda$ that satisfies $w^2 \le c (1-\Delta)$ as an operator inequality on $L^2(\Lambda )$ for some constant $ c \in [0,1)$.
\item[{\color{white}{aaa}}\hypertarget{(A$v_L$)}{(A$v_L$)}] There is a rotational invariant function $\widehat v_\infty(k) : \mathbb R^d \to \mathbb R$  (not depending on $L$) with
\begin{align}
\vert \widehat v_\infty(k) \vert\, \le\, (k ^2 +  R )^{-1}
\end{align}
for some $R>0$ and $\widehat v_L (k) =  \widehat v_\infty (k)$ for all $k \in (2\pi /L) \mathbb Z^d$.\footnote{We use the convention
\begin{align}
\widehat v_L(k) = \int_\Lambda \D^d x\, v_L(x) e^{-ikx}
\end{align}
} 
\end{itemize}
It is well known that, under these conditions, 
%$w$ and $v_L$ are infinitessimally bounded with respect to the Laplace operator, meaning that for any $\varepsilon>0$ there exists a $C_\varepsilon>0$ such that $\sno v_L \psi \sno \le \varepsilon \sno \Delta \psi \sno + C_\varepsilon$ for all wave functions $\psi\in L^2(\Lambda)$ (and the same for $w$). By the Kato-Rellich theorem this implies that %
$H$ defines a self-adjoint operator.

In Proposition \ref{prop: lower bound for h_0} we also require that $\vert \widehat v_\infty (k)\vert \ge ( 1 + R )^{-1}$ for all $ k^2 \le 1$. Hence it is suggestive to think of $\widehat v_\infty (k) =  ( k^2 + R)^{-1}$, which, up to some constants, is the Fourier transform of a Yukawa potential.

In our main results we choose the coupling constant $\lambda$ proportional to $\varrho^{(2-d)/2d}$ with $\varrho = N L^{-d}$, and then analyze the regime $\varrho \gg 1$. To be more precise, we will first take the large-volume limit $L\to \infty$ with $\varrho>0$ constant and then consider $\varrho \gg 1$. As will be explained in Section \ref{sec: effective model}, the scaling of $\lambda$ is chosen such that we have a non-trivial effective dynamics.

Our goal is to analyze the solution of the time-dependent Schr\"odinger equation
\begin{align}
\begin{cases} i \frac{d}{dt}  \Psi(t) &\hspace{-3mm} = H \Psi(t) \\[0mm]
\hspace{4.75mm} \Psi(0) & \hspace{-3mm} = \Psi_0 
\end{cases} 
\end{align}
for initial states $\Psi_0 \in \mathscr H_n \otimes \mathscr H_N^-$ of the form
\begin{align}\label{eq: initial state}
\boxed{\quad 
 \Psi_0 (y_1 ,\ldots , y_n , x_1,...,x_N) = \xi_0(y_1,\ldots, y_n) \otimes \Omega_0 (x_1,\ldots ,x_N) .\quad 
}
\end{align}
We assume that the fermions are initially in the ground state of the non-interacting Fermi gas (the Fermi sea), that is, the ground state of the kinetic energy operator $T_N$. The $n$-body wave function $\xi_0 \in \mathscr H_n$ can be chosen more generally. Our only requirement is that its kinetic energy is of order one with respect to $\varrho \gg 1$, by which we ensure an important separation of scales between the impurity particles and the fast fermions. No statistics are imposed on the impurities.

Instead of having $N$ as a free model parameter, it is more convenient to choose a Fermi momentum $k_{F}>0$, and then fix $N$ in terms of $k_F$ and $L$ by
\begin{align}\label{eq: Fermi ball}
 N = N(k_F , L ) = \vert B_F \vert  \quad \text{with}\quad   B_F = \big\{ k \in (2\pi /L) \mathbb Z^d\, : \, \vert k \vert \le k_{ F} \big\} .
\end{align}
This implies the non-degeneracy of the free fermionic ground state, given by the anti-symmetric product of all plane waves with momenta inside the Fermi ball $B_F$,
\begin{align}\label{def: Fermi sea}
\Omega_0 = \bigwedge_{k \in B_F} \varphi_k \in \mathscr H_{N(k_F,L)}^- , \quad  \varphi_k(x ) = \frac{ \exp(i k x) }{ L^{d/2}}\in L^2(\Lambda).
\end{align}
Clearly, $(T_N -E^0(k_F,L) ) \Omega_0 = 0$ with eigenvalue $E^0( k_{ F} ,L) = \sum_{k\in B_F} k^2$.

Replacing the sum by its Riemann integral, one obtains the useful relation between the Fermi momentum and the average density,
\begin{align}\label{eq: rel rho and kF}
 \frac{ N( k_{ F} ,L) }{L^d} =    V_d  \, k_{ F}^d + o (1) 
\end{align}
where $o(1)$ vanishes as $L\to \infty$, and the constants equal $V_1 =  1/\pi$, $V_2 = 1/(4\pi)$ and $V_3 = 1/(6\pi^2)$. From \eqref{eq: rel rho and kF} one infers that $\varrho \gg 1$ is equivalent to $k_F\gg 1$.

\subsection{Effective $n$-body model} \label{sec: effective model}

For non-vanishing interaction between the impurities and the fermions (i.e., for $v_L \neq 0$), one can not expect that the time-evolved wave function $\Psi(t)= e^{-iHt} \Psi_0$ exhibits the same product form as the initial state $\Psi_0 = \xi_0 \otimes \Omega_0$. Nevertheless, by including an effective interaction among the impurities, we shall show that the product structure is approximately preserved in the limit of large $k_F$. To this end, we compare $\Psi(t)$ with the product wave function
\begin{align}\label{eq: effective dynamics}
\Psi^{\rm{eff}}(t) = e^{-i h_n  t }  \xi_0 \otimes  e^{-i  E(k_F,L)t} \Omega_{0}
\end{align}
where $E(k_F,L)$ is the energy shift
\begin{align}\label{eq: energy shift}
E(k_F,L) = E^0(k_F ,L) + n \lambda  \widehat v_L(0) \frac{N(k_F,L)}{L^{d}},
\end{align}
and $h_n$ the an operator on $\mathscr H_n$ defined by
\begin{align}\label{eq: effective n body hamiltonian}
\begin{boxed}
{
\quad h_n =   h^0_n -  \sum_{i<j }^n  \lambda^2  W_{k_F } ( \vert y_i - y_j \vert )   - n  \lambda^2 W_{k_F } (0)  .\quad 
}
\end{boxed}
\end{align}
Here, we introduced the effective interaction potential
\begin{align}\label{eq: effective potential}
 W_{k_F }( r ) =   V_d^2 \int\limits_{\vert k \vert \le k_F } \D^d k \int\limits_{\vert l \vert > k_F } \D^d l  \ \frac{\vert \widehat v_\infty (l-k) \vert^2}{ l^2 - k^2 + (l-k)^2 + 1  }  \cos((l -k )\cdot r \hat a  )
\end{align}
for $r\ge 0$ and $\hat a \in \mathbb R^d$ an arbitrary vector of unit length (since $\widehat v_\infty$ is rotational invariant the direction of $\hat a$ is irrelevant).
%The plus one in the denominator is added for convenience since replacing the denominator by $l^2-k^2 + (l-k)^2$ would only lead to a subleading effect. 
In Lemma \ref{lem: lower bound effective potential} below we show that $W_{k_F}(r)$ is a bounded function and thus $h_n$ is self-adjoint and generates the unitary time evolution $e^{-i h_n  t}$.

Before we discuss $W_{k_F}(r)$ in more detail, let us comment on the physical interpretation of the effective dynamics defined by \eqref{eq: effective dynamics}:
\begin{itemize}
\item[(a)] There is no interaction between the impurities and the fermions.\\[-8mm]
\item[(b)] The time-evolution of the fermions is stationary.\\[-8mm]
\item[(c)] The impurities evolve with an additional pair interaction described by the potential $- \lambda^2 W_{k_F } ( r )$. (The last term in \eqref{eq: effective n body hamiltonian} only adds a constant phase shift.)
\end{itemize}
The heuristic picture behind the effective interaction is that it is caused by particle-hole excitations in the Fermi sea. One of the impurities produces a particle-hole excitation in the Fermi sea and then a different impurity annihilates the particle-hole excitation again. After such a second-order process the impurity particles are obviously correlated with each other but not with the Fermi sea.\footnote{Similar processes can of course involve more than two impurities which would lead to more complicated effective interactions. In our setting, however, these would be of subleading order and thus need not be taken into account in the effective dynamics.} We remark that this mechanism is in principle similar to the effect of vacuum polarization in QED if one interprets the Fermi sea as the vacuum.
%Analogous effects were discussed in the physics literature also for systems of ultra cold atoms, see for instance \cite{EnsEtAl2020,Huang20,Kinnunen15,MistakidisEtAl18,Nishida2009,Santamore08}.

Since we are interested in the limit of large density, it is important to understand the properties of the effective potential for large values of $k_F$. In our main results we choose the coupling constant $\lambda $ such that
\begin{align}
\begin{boxed}
{\quad  \lambda^2 =  k_F^{(2-d)} . \quad }
\end{boxed}
\end{align}
This is motivated by the next lemma stating that for such $\lambda$, the effective potential $\lambda^2 W_{k_F}(r)$ is of order one with respect to $k_F\gg 1$. In Proposition \ref{prop: lower bound for h_0} we shall use this property to show that the effective interaction can not be omitted in the effective dynamics.

\begin{lemma}\label{lem: lower bound effective potential}
Let $W_{k_F}(r)$ be defined by \eqref{eq: effective potential} with $\vert \widehat v_\infty(k) \vert \le (k ^2 + R )^{-1}$ for all $k\in \mathbb R^d$ and some $R>0$, and $\vert \widehat v_\infty (k) \vert \ge (1 + R )^{-1}$ for all $k^2 \le 1$. It follows that there are constants $C>1$ and $c>0$ such that
 \begin{align}\label{eq: lower bound effective potential}
\sup_{k_F\ge 1} \, \sup_{r \in [0,\infty)} \, \vert k_F^{(2-d)} W_{k_F}(r) \vert \,  \le\, C  \quad \text{and} \quad  \inf_{k_F\ge 1 } \, \inf_{r\in [0,c]} \,  k_F^{(2-d)} W_{k_F}(r)   \, \ge \,  C^{-1}.
\end{align}
%Moreover $k_F^{(2-d)}  W_{k_F}(r)\to 0 $ as $r\to \infty$.
\end{lemma}

Below we visualize these qualitative properties by showing a numerical computation for $k_F^{(2-d)}W_{k_F}(r)$ as a function of $r\ge 0$. We set $d=2$ and $\widehat v_\infty (k) = 1/2$ for $k^2 \le 1$ and zero otherwise. The graph is plotted in $k_F$-independent units of length (x-axis) and energy (y-axis). Since the single points converge rapidly for growing values of $k_F$ we only depict it for one value. The picture is qualitatively the same in other dimensions.

\begin{figure}[!htb]
        \center{ \includegraphics[width=0.75\textwidth]
        {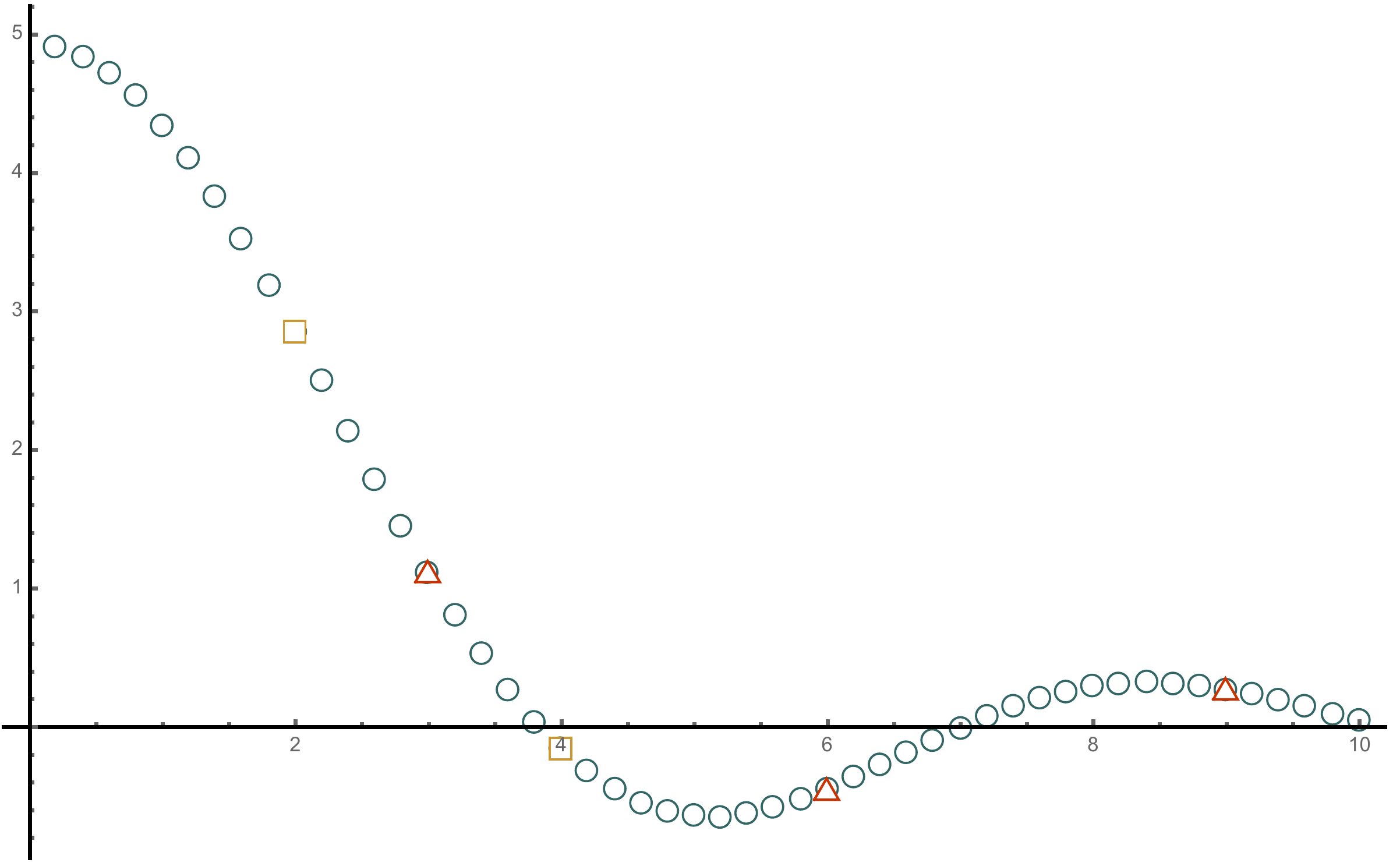}}
        \caption{\label{fig:my-label} Effective pair potential $k_F^{(2-d)}W_{k_F}(r)$ in $k_F$-independent units\label{figure 1}} 
\end{figure}

\subsection{Main results}\label{sec: main results}

We are now ready to state our main theorem which provides an estimate for the large-volume limit of the norm of $\Psi(t) - \Psi^{\rm eff}(t) $. For reasons explained in the remarks, we only consider $d\in \{2,3\}$. Note that we indicate the norm resp. the scalar product on the Hilbert space $\mathscr H_n \otimes \mathscr H_N^{-}$ by $\sno \cdot \sno$ and $\small \langle \cdot, \cdot \small \rangle$ whereas for the spaces $ \mathscr H_n$ and $ \mathscr H_N^-$, we add additional subscripts.

\begin{theorem} \label{theorem: main theorem} Let $d\in \{2,3\}$, $n\ge 2$ and assume \textnormal{(}\hyperlink{(A$w$)}{\textnormal{A}$w$}\textnormal{)} and \textnormal{(}\hyperlink{(A$v_L$)}{\textnormal{A}$v_L$}\textnormal{)}. For $L ,k_F >0$ choose $N=N(k_F,L)$ as in \eqref{eq: Fermi ball} and let $\Omega_0$ be given by \eqref{def: Fermi sea}. Let further $\xi_0 \in \mathscr H_n$ obey the conditions $\sno \xi_0 \sno_{\mathscr H_n}  = 1$ and 
\begin{align}\label{eq: condition on the kinetic energy}
q_{\xi_0} := \sup_{k_F\ge 1} \sup_{ L>0} \sum_{i=1}^n \lsp \xi_0,  (- \Delta_{y_i})  \xi_0 \rsp_{\mathscr H_n}   <\infty,
\end{align}
and let $E(k_F ,L)$ and $h_n$ be defined by \eqref{eq: energy shift} and \eqref{eq: effective n body hamiltonian}, respectively. For $\vert \lambda \vert =  k_F^{(2-d)/2}$ the wave functions $\Psi(t)  = e^{-iHt} \xi_0 \otimes \Omega_0$ and $\xi(t) = e^{-ih _n t} \xi_0$ satisfy the following property. There exists a constant $C(n,R, q_{\xi_0}) >0$ such that
\begin{align}\label{eq: main estimate}
\boxed{\quad 
\limsup_{ L \to \infty } \sno   \Psi(t)   - \xi(t) \otimes  e^{-i E (k_F,L)  t}  \Omega_0  \sno  \le C(n,R, q_{\xi_0}) \,  \frac{ (1+ \vert t \vert )  ( \ln  k_F)^{3}}{ \sqrt { k_F } } \quad }
\end{align}
for all $k_F\ge  2 $ and $t\in \mathbb R$.
\end{theorem}

Since all wave functions on the left side of \eqref{eq: main estimate} are normalized to one, the bound is meaningful if the right side is small compared to one. This is the case for $k_F \gg 1 $ as long as $\vert t \vert\ll  \sqrt{ k_F }  (  \ln  k_F )^{-3}$.\\

\noindent \textbf{Remarks}. \noindent 1.1. As a simple corollary one gets convergence of the reduced densities (in trace norm distance), $\gamma^{(n)}_{\Psi(t)} = \textnormal{Tr}_{\mathscr H_N^-}  \vert \Psi(t) \rangle \langle \Psi(t) \vert $ and $\mu^{(N)}_{\Psi(t)} = \textnormal{Tr}_{\mathscr H_n}  \vert \Psi(t) \rangle \langle \Psi(t) \vert $ towards $\vert \xi(t) \rangle \langle \xi(t) \vert $ and $\vert \Omega_0 \rangle \langle \Omega_0 \vert$, respectively:
\begin{align}\label{eq: main estimate reduced density}
\limsup_{ L \to \infty } & \bigg( \textnormal{Tr}_{\mathscr H_n} \Big\vert \gamma^{(n)}_{\Psi(t)} - \vert \xi(t) \rangle \langle \xi(t) \vert \Big\vert  + \textnormal{Tr}_{\mathscr H_N^-} \Big\vert \mu^{(N)}_{\Psi(t)} - \vert \Omega_0  \rangle \langle \Omega_0 \vert \Big\vert  \bigg)  \notag \\
& \hspace{5.5cm} \le C(n,R, q_{\xi_0}) \,  \frac{ (1+ \vert t \vert )  ( \ln  k_F)^{3}}{ \sqrt { k_F } }.
\end{align}
%This follows from the inequality 
%\begin{align}\label{eq: trace inequality}
%\textnormal{Tr}_{\mathscr H_A} \big\vert \textnormal{Tr}_{\mathscr H_B} \vert \Psi \rangle \langle \Phi \vert \big\vert \le \sno  \Psi \sno_{\mathscr H_A \otimes \mathscr H_B}  \, \sno \Phi \sno_{\mathscr H_A \otimes \mathscr H_B}
%\end{align}
%which is valid for $\Psi,\Phi\in \mathscr H_A \otimes \mathscr H_B $ where $\mathscr H_A$ and $\mathscr H_B$ are two separable Hilbert spaces. For a short proof of \eqref{eq: trace inequality} we refer to \cite[Appendix D]{FrankG2017}.\medskip

\noindent \hypertarget{rem: coupling parameter}{1.2.} Our proof provides a more general statement than Theorem \ref{theorem: main theorem}. We shall show that for $d\in \{1,2,3\}$, $\lambda \in \mathbb R$ and all $k_F\ge 2$, it holds that 
\begin{align}
\label{eq: main estimate remark}
& \limsup_{ L \to \infty } \sno   \Psi(t)   - \xi(t) \otimes  e^{-i E (k_F,L) t}  \Omega_0  \sno  \le C(n,R,q_{\xi_0})\,  \Gamma (d,k_F,\lambda , t ) 
\end{align}
with \allowdisplaybreaks
\begin{align} \label{def: GAMMA(d)}
\Gamma (d,k_F,\lambda,t) & = \vert  \lambda  \vert \big(  1   + \vert t \vert  (1 + \lambda^2 k_F^{(d-2)} )   \big)   k_F^{(d-3)/2}  (\ln k_F)^{1/2}    \notag \\[2.5mm]
&\quad+   \lambda^2  \big(   k_F^{(d-3)}  \ln k_F        + \vert t \vert  (1+\lambda^2 k_F^{(d-2)})  k_F^{(d-3) } \ln k_F +   \vert t\vert  \sqrt{\gamma(d,k_F)}   \big)      \notag \\[2.5mm] 
& \quad  +  \vert \lambda\vert ^3 \vert t \vert \big(    k_F ^{ ( 3d-7 ) / 2 }  \ln k_F  +  \sqrt{ \gamma (d,k_F)} k_F^{(d-3)/2} (\ln k_F)^{1/2} + \gamma (d,k_F) \big)
\end{align}
and
\begin{align}\label{def: Gamma(d)}
\gamma (d,k_F) = \begin{cases} k_F^{(2d-5)} (\ln k_F)^3 \quad \textnormal{for}\ d=2,3\\[1mm]
k_F^{-2} (\ln k_F)^3 \quad \quad \ \textnormal{for}\ d=1 \end{cases}.
\end{align}
Setting $\vert \lambda \vert = k_F^{(2-d)/2}$ leads to \eqref{eq: main estimate}.\medskip

\noindent 1.3. For $d=1$ the natural choice of the coupling constant (such that the effective interaction is of order one) is $\lambda^2 = k_F$. In this case, however, the error term $\lambda^2 \sqrt{ \gamma(d,k_F)} $ is not small, and thus the right side in \eqref{eq: main estimate remark} does not provide a useful bound.\footnote{For $\lambda = o(k_F (\ln k_F)^{-3/2}) $, the upper bound is still useful, but this case is less interesting since the effective potential would be of subleading order and, in particular, of the same order as the error.} Nonetheless, we expect that a modification of the proof would allow the derivation of Theorem \ref{theorem: main theorem} also for $d=1$ and $\lambda^2 = k_F$. Since we prefer to keep the presentation at a considerable length, we omit this case and restrict our analysis to $d\in \{2,3\}$.\medskip

\noindent 1.4. Theorem \ref{theorem: main theorem} is valid also for $n=1$ if one uses $h_{n=1} = -\Delta_y - \lambda^2 W_{k_F}(0)$. This improves our findings in \cite{JeblickMP2018}.\medskip

Our second result shows that the effective interaction in \eqref{eq: effective n body hamiltonian} adds a non-negligible effect to the dynamics. In other words, a result like Theorem \ref{theorem: main theorem} can not be true if one replaces the $n$-body Hamiltonian $h_n$ by 
\begin{align}\label{eq: h tilde}
\widetilde h_n =  h^0_n -  n  k_F^{(2-d)}  W_{k_F}(0).
\end{align}
 
\begin{proposition}\label{prop: lower bound for h_0} Assume the same conditions as in Theorem \ref{theorem: main theorem}, and in addition, let $\vert \widehat v_\infty(k)\vert \ge 1 / (1 + R )  $ for all $k^2 \le 1$ and assume
\begin{align}\label{eq: initial states concentrated at y_1 equal y_2}
\inf_{ k_F\ge 1  } \inf_{ L >0 } \, \sum_{i<j}^n k_F^{(2-d)}  \lsp \xi_0, W_{k_F}(\vert y_i-y_j \vert ) \xi_0\rsp_{\mathscr H_n}  \ge c_0 \quad \text{and} \quad \sup_{ k_F\ge 1  } \sup_{ L >0 } \, \sno h^0_n \xi_0 \sno_{\mathscr H_n} \le C_0
\end{align}
for some constants $c_0, C_0 >0$. Then there exist times $t_1> t_0 >0$ \textnormal{(}depending on $c_0,C_0$, $n$, $R$ and $q_{\xi_0}$\textnormal{)} such that
\begin{align}\label{eq: dynamics lower bound}
\boxed{
\quad  \liminf_{ L \to \infty } \sno   \Psi(t )   -  \exp( -i  \widetilde h_n  t ) \xi_0  \otimes  e^{-i  E (k_F,L) t } \Omega_0 \sno  \ge  \frac{ c_0 }{2} t  \quad
}  
\end{align}
for all $t \in (t_0,t_1)$ and all $k_F$ large.
\end{proposition}

We emphasize that assumptions \eqref{eq: condition on the kinetic energy} and \eqref{eq: initial states concentrated at y_1 equal y_2} are not very restrictive and, in particular, the three conditions are consistent with each other. This is thanks to Lemma \ref{lem: lower bound effective potential} that allows us to locate some mass of $\xi_0$ inside the non-vanishing positive core of the effective potential while keeping the values of $\sno h^0_n \xi_0 \sno$ and $q_{\xi_0}$ of order one as $k_F$ tends to infinity.\medskip

We conclude this section with a short sketch of the strategy behind the proof of Theorem \ref{theorem: main theorem}. The starting point is to use the fundamental theorem of calculus to write
\begin{align} 
&  \big( 1 -  e^{ i (  H-E(k_F,L)  ) t}  e^{-i h_n t}  \big) \xi_0 \otimes  \Omega_{0}  \notag\\
& \hspace{1cm} =  - i \int_0^t \D s\,  e^{i (H-E(k_F,L)) s} \bigg( \underbrace{ V_{N+n}  - n \lambda \widehat v_L (0) \frac{N(k_F,L)}{L^d} }_{=:V_{N+n}^{(\rm {exc})}}+ h_n^0 - h_n \bigg)   \xi(s) \otimes \Omega_0 . \label{eq: sketch of the proof}  
\end{align}
To estimate the norm of the part with $V_{N+n}^{(\rm {exc})}$, we need to use the unitary $e^{i (H-E(k_F,L))s}$. The heuristic idea is that the operator $ T_N^{(\text {exc})} := T_N-E^0(k_F,L) $ in 
\begin{align}
H-E(k_F,L) =  T_N^{(\text {exc})}  + h^0  + V_{N+n}^{(\text{exc})} 
\end{align}
produces a large energy shift when applied to states orthogonal to $\Omega_0$. This, in turn, leads to phase cancellations (or destructive interference) in \eqref{eq: sketch of the proof} and thus suppresses the value of the norm. The obvious way to exploit such phase cancellations is to employ the identity
\begin{align}\label{eq: phase cancellations}
i\, e^{i (H-E(k_F,L)) s}  = e^{i (H-E(k_F,L))s} e^{-i T_N^{(\text {ex})}  s} \Big( \frac{d}{ds} e^{i  T_N^{( \text {exc}) }  s} \Big) (T_N^{(\text {exc}) })^{-1} 
\end{align}
and then use integration by parts (note that $V_{N+n}^{(\rm {exc})}\Omega_0$ is orthogonal to $\Omega_0$ and thus \eqref{eq: phase cancellations} can be applied). This leads to a perturbation type expansion for the first part in \eqref{eq: sketch of the proof} that involves terms with expressions like $V_{N+n}^{({\rm exc})}( T_N^{( \text {exc} ) } )^{-1}V_{N+n}^{({\rm exc})} \xi(s) \otimes \Omega_0$. However, since not all of the terms in this expansion are sufficiently small, we we need to use the unitary again and proceed by a second integration by parts. To do that, we now have to sort the terms into a component along $\Omega_0$ and all other components orthogonal to $\Omega_0$. In the first component there are no more phase cancellations since $T_N^{( \text {exc} ) } \Omega_0 =0$ (the points of stationary phase so to say). This part is canceled by the second term in \eqref{eq: sketch of the proof}, which follows from
\begin{align}
h_n = h_n^0 - \lsp \Omega_0, V_{N+n}^{({\rm exc})}( T_N^{( \text {exc} ) } )^{-1}V_{N+n}^{({\rm exc})} \Omega_0 \rsp_{\mathscr H_N^-},
\end{align}
and it is this cancellation that determines the choice of $h_n$. For all other terms, the ones orthogonal to $\Omega_0$, we can proceed by a second expansion via integration by parts. All terms that are obtained by this expansion are then estimated separately.

As a final remark let us mention that a similar strategy was recently used in \cite{Leopold2020,Mitrouskas2020} to study the quantum fluctuations of the dynamics of a strongly coupled polaron. In this model, the term without the oscillating phase leads to an effective quadratic interaction between the phonons inside the polaron cloud.

\section{Proofs \label{sec: proofs}}

We first introduce the formalism of second quantization and then state some preliminary estimates for sums of different transition amplitudes. The bounds for the transition amplitudes are required throughout the proof of Theorem \ref{theorem: main theorem}. (To shorten the presentation of this section, we provide some more bounds and all proofs in Appendix \hyperlink{A}{A}.)

\subsection{Second quantization}

\noindent We think of the Hilbert space $\mathscr H_n \otimes \mathscr H_N^-$  as the $N$-particle sector of $\mathscr H_n \otimes  \mathcal F$ with $\mathcal F$ the fermionic Fock space
\begin{align}
\mathcal F = \bigoplus_{m=0}^\infty \mathscr H_m^- , \quad  \mathscr H_m^- = \bigwedge^m L^2(\Lambda).
\end{align}
This way we can profit from the use of the formalism of second quantization, which strictly speaking is not necessary, but simplifies many computations.

For plane waves $\varphi_k(x) = L^{-d/2}\exp(ikx) $, $k\in (2\pi /L )\mathbb Z^d $, we define the creation and annihilation operators $a_k ,a^*_k: \mathscr H_n \otimes \mathcal F  \to  \mathscr H_n\otimes \mathcal F$ through
\allowdisplaybreaks
\begin{align}
 (a_k \Psi)^{(m)}(y_1,...y_n,x_1,...,x_m) &  = \sqrt{m+1} \int_{\Lambda } \D x\, \overline{\varphi_k(x)} \Psi^{(m+1)}(y_1,...y_n,x_1,...,x_m,x),   \notag \\
(a^*_k \Psi)^{(m)}(y_1,...,y_n,x_1,...,x_m) &  = \sum_{j=1}^m\frac{(-1)^j}{\sqrt m}\, \varphi_k(x_j) \Psi^{(n-1)}(y_1,...,y_n ,x_1,..., x_{j-1},x_{j+1} ,...,x_m), \notag  
\end{align}
where $\Psi = (\Psi^{(m)})_{m\ge 0}$ with $\Psi^{(m)}\in \mathscr H_n \otimes \mathscr H_m^-$. They satisfy the usual canonical anti-commutation relations
\begin{align}
 a^*_k a_l  +  a_l a_k^*    = \delta_{kl}, \quad  a_k  a_l + a_l a_k   = 0\quad \forall k,l \in (2\pi /L )\mathbb Z^d.
\end{align}

In terms of creation and annihilation operators we can write the Hamiltonian $H$ as an operator on the Hilbert space $\mathscr H_n \otimes \mathcal F$. For that purpose, define
\begin{align}\label{eq: second quantized Hamiltonian}
\mathbb H  =  h^0_n + T  + \mathbb V - E^0 (k_F,L) 
\end{align}
with 
\begin{align} 
T=  \sum_{k}k^2 a_ k^* a_k, \quad  \mathbb V =  \sum_{i=1}^n \mathbb V^{(i)}  , \quad   \mathbb V^{(i)} = \lambda L^{-d}\sum_{k\neq l} \widehat v_L (l-k)   e^{i(k-l) y_i }   a_l^* a_k.
\end{align}
Restricting $\mathbb H$ to the $N$-particle sector yields
\begin{align}\label{eq: restriction of H}
\mathbb H \restriction  \mathscr H_n \otimes \mathscr H_{N(k_F,L)}^-  = H - E(k_F,L)
\end{align}
with $E(k_F,L)$ defined in \eqref{eq: energy shift}. Note that for ease of notation we subtracted the energy $E^0(k_F,L)$ and also the momentum-conserving part of the interaction
\begin{align}
\lambda \sum_{i=1}^n   L^{-d}\sum_{k,l} \widehat v_L (l-k)    e^{i( k-l ) y_i  } \delta_{kl} \, a_l^* a_k = \frac{n \lambda \widehat v_L (0)}{L^d} \sum_{k} a_k^*a_k.
\end{align}

In terms of the Fock space vacuum $\vert 0 \rangle = (1,0,0,\ldots)$, the Fermi sea \eqref{def: Fermi sea} is given by $\Omega_0 = ( \prod_{k\in B_F} a_k^* ) \vert 0 \rangle $. Because of the anti-commutation relations, it satisfies
\begin{align}\label{eq: Fermi sea properties}
\begin{cases}
a_k \Omega_0 & = \quad 0\quad \text{for all}\ k\in B_F^c,\\[1mm]
a^*_k \Omega_0 & = \quad 0\quad \text{for all}\ k\in B_F.
\end{cases}
\end{align}
Moreover, for momenta $l'_1,...,l'_n,l_1....,l_n\in B_F^c$ and $k'_1,...,k'_m,k_1,...,k_m\in B_F$ (with integers $n+ m\ge 1$) the anti-commutation relations together with \eqref{eq: Fermi sea properties} imply the Wick formula
\begin{align}\label{eq: Wick formula}
& \lsp a_{l'_1}^* \dots a_{l'_n}^* a_{k'_1} \dots a_{k'_m} \Omega_{0}, a_{l_1}^* \dots a_{l_n}^* a_{k_1} \dots a_{k_m}  \Omega_{0} \rsp_{\mathscr H_N^-} \nonumber \\
& \hspace{4cm} = \Big( \sum_{\sigma \in S_n} \text{sgn}(\sigma) \prod_{i=1}^n \delta_{l_i l_{\sigma(i)}'}  \Big)  \Big( \sum_{\tau \in S_m} \text{sgn}(\tau)  \prod_{j=1}^m \delta_{k_j k_{\tau(j)}'} \Big) ,
\end{align}
where $\text{sgn}(\sigma) \in \{-1,1\}$ is the sign of the permutation $\sigma \in S_n$ (the symmetric group).

\subsection{Preliminary bounds}

\noindent Throughout this section we use the notation 
\begin{align}
\sum_{(l,k)\in A} f(l,k) = \sum_k \sum_l \chi_A(l,k) f(l,k),
\end{align}
where $\chi_A$ denotes the characteristic function $\chi_A(l,k) = 1$ for $(l,k) \in A$ and $\chi_A(l,k) = 0$ otherwise. To state the next lemma, let us introduce the set of momentum pairs
\begin{align}
T_F  =   \big\{ (l,k) \in B_F^c\times B_F   \big\}\subset ( 2\pi /  L)  \mathbb Z^d\times (  2\pi / L ) \mathbb Z^d. \label{def: momentum pairs TF} 
\end{align}

\begin{lemma}\label{lem: a priori bounds} Let $d\in \{1,2,3\}$ and assume \textnormal{(}\hyperlink{(A$v_L$)}{\textnormal{A}$v_L$}\textnormal{)}. There is a constant $C>0 $ \textnormal{(}depending on $R$\textnormal{)} such that for all $k_F\ge 2$ the following estimates hold.
\allowdisplaybreaks
\begin{subequations}
\begin{align}
 \lim_{  L\to\infty  }   \Bigg( L^{-2d} \sum_{(l,k) \in T_F} \vert \widehat v_L (l - k) \vert^2    \Bigg)  & \le C k_F^{(d-1)}\label{eq: sum of transitions 1},
\\[1mm]
\lim_{  L\to\infty  } \Bigg(  L^{-2d} \sum_{(l,k) \in T_F} \frac{ \vert \widehat v_L (l - k) \vert^2   }{l^2-k^2 + 1 }  \Bigg) & \le C k_F^{(d-2)} \label{eq: sum of transitions 2},\\[1mm]
\lim_{ L\to\infty }  \Bigg( L^{-2d}\sum_{(l,k)\in T_F} \frac{ \vert \widehat v_L  (l - k) \vert^2  }{(l^2-k^2+1 )^2 }\Bigg) & \le  C    k_F^{(d-3)} \ln k_F  ,\label{eq: sum of transitions 3}\\[1mm]
\lim_{ L\to\infty }  \Bigg( L^{-2d}\sum_{(l,k)\in T_F} \frac{ \vert \widehat v_L (l - k) \vert^2 ( l-k )^2 }{(l^2-k^2+(l-k)^2 + 1 )^2 }\Bigg) & \le  C k_F^{(d-3)}   (\ln k_F)^2  . \label{eq: sum of transitions 4}
\end{align}
\end{subequations}
\end{lemma}

These bounds will be frequently used in the next section. Their derivation is postponed to Appendix \hyperlink{A}{A}. Some more bounds, of similar type, are provided in Lemma \ref{lem: a priori bounds 2}.

\subsection{Proof of Theorem \ref{theorem: main theorem}}

In this section we derive the bound stated in Remark \hyperlink{rem: coupling parameter}{1.2}, that is, we keep $\lambda \in \mathbb R$ arbitrary and consider $d\in \{1,2,3\}$. Theorem \ref{theorem: main theorem} is a consequence of this bound for $\vert \lambda \vert = k_F^{(2-d)/2}$. 

Our goal is to estimate the norm difference
\begin{align}
\sno e^{-i (H-E) t } \xi_0 \otimes \Omega_{0} -  \xi(t) \otimes  \Omega_{0} \sno
\end{align}
with $\xi(t) = e^{-ih_n t}\xi_0$. For shorter notation, we omit from now on the arguments in
\begin{align}
E= E(k_F,L), \quad  E^0= E^0(k_F,L) \quad \text{and}\quad  N = N(k_F,L).
\end{align}
To start we employ \eqref{eq: restriction of H} and use the fundamental theorem of calculus to get
\begin{align}\label{eq: psi minus psi ph}
&  \big( 1 -  e^{ i \mathbb H  t}  e^{-ih_n t } \big) \xi_0 \otimes  \Omega_{0}  =  - \int_0^t \D s \, \frac{d}{ds}  \big( e^{ i \mathbb H s}  e^{-ih_n s } \big)  \xi_0 \otimes  \Omega_{0}   =: \Phi(t) + \phi(t)
\end{align}
with
\begin{align}
\Phi (t)&  = 
-i \int_0^t \D s \
e^{  i  \mathbb  H   s}   \mathbb V  \xi(s) \otimes \Omega_{0} \quad  \phi(t) =  i \int_0^t \D s \
e^{  i  \mathbb  H  s}   (h_n -h_n ^0) \xi(s) \otimes \Omega_{0}.
\end{align}

In the first part of the proof we use two integration by parts in order to expand the state $\Phi(t)$ into several contributions. In the second part we estimate these contributions separately. In particular we single out one contribution which is canceled by $\phi(t)$.

\paragraph{Decomposition of $\Phi(t)$} {\textcolor{white}{a}}\medskip

We define the resolvent type operator 
\begin{align}
R = (T -  E^0 + P_{\rm f}^2 + 1 ) ^{-1} \restriction \mathscr H_n \otimes \mathscr H_N^-
\end{align}
with $P_{\rm f} = \sum_{k} k  a^*_k a_k$ the momentum operator of the fermions. Since $T \restriction \mathscr H_n \otimes \mathscr H_N^- \ge E^0$, the operator $R$ is bounded. Here we deviate slightly from the strategy explained at the end of Section \ref{sec: main results}, where we used $(T-E^0)^{-1}$ instead of $R$. While the plus one is added in order to avoid a potential singularity from contributions whose energy excitation vanishes in the limit $L\to \infty$, the momentum part $P_{\rm f}^2$ needs to be included for important cancellations.\footnote{\label{footnote: pf squared}As we allow for singular interaction potentials $v(x)$, the gain in kinetic energy of the impurity after interacting with a fermion can be very large. Adding the operator $P_{\rm f}^2$ in $R$ will lead to a cancellation of this gain in kinetic energy, cf. \eqref{eq: Phi h term}.}

With the aid of $R$ we can rewrite
\begin{align}\label{eq: Phi with resolvent}
\Phi(t) & = -    \int_0^t \D s\, e^{i  \mathbb H   s} e^{-i (T - E^0 + 1 + P_{\rm f}^2) s}  \Big( \frac{d}{ds} e^{i (T - E^0 +1  + P_{\rm f}^2 ) s} \Big) R  \mathbb V   \xi(s) \otimes \Omega_0 
\end{align}
and integrate by parts. This leads to $\Phi(t) = \sum_{i=0}^3 \Phi_{i}(t) $ with\allowdisplaybreaks 
\begin{subequations}
\begin{align}
\Phi_{0}(t) & =   - \Big[ e^{i  \mathbb H   s}    R \mathbb V   \xi(s) \otimes \Omega_0  \Big]_0^t  , \label{eq: Phi boundary term}\\[0.5mm]
\Phi_{1}(t) & =   i \int_0^t \D s\, e^{i  \mathbb H  s}  R \mathbb V  (-1  + h_n^0 - h_n )   \xi(s) \otimes \Omega_0 ,  \label{eq: Phi boundary term 2 }\\[0.5mm]
\Phi_{2}(t)  &  = i   \int_0^t \D s\, e^{i  \mathbb H  s} R \big\{ [ h^0_n , \mathbb V ]  - P_{\rm f}^2  \mathbb V \big\} \xi(s) \otimes \Omega_0  ,
\label{eq: Phi h term}\\[0.5mm]
\Phi_3(t) & = i   \int_0^t \D s\, e^{i  \mathbb  H  s } \mathbb V  R  \mathbb V  \xi(s) \otimes \Omega_0 , \label{eq: Phi Interaction term}
\end{align}
\end{subequations}
where we used $\frac{d}{ds}  \xi(s)  = -i h_n \xi(s)$ and
\begin{align}
\frac{d}{ds}  e^{i  \mathbb H   s} e^{-i (T - E^0 + 1 + P_{\rm f}^2) s}   & = i e^{i  \mathbb H  s}  \big( h^0_n   + \mathbb V  - 1  - P_{\rm f}^2 \big)  e^{-i (T -E^0 + 1 + P_{\rm f}^2) s}.
\end{align}

In line \eqref{eq: Phi Interaction term} we proceed by decomposing the state $\mathbb V R \mathbb V \xi(s) \otimes \Omega_0$ according to the number of holes in the Fermi sea $\Omega_0$. (A hole is a an unoccupied momentum mode with $k \in B_F$; since we only consider states in $\mathscr H_N^-$, a state with $m$ holes is automatically a state with $m$ particle-hole pairs where particle refers to an occupied mode with $k \in B_F^c$). We use that the operator $\mathbb V R \mathbb V$ changes the number of holes at most by two, and thus the state $\mathbb V R \mathbb V \xi(s) \otimes \Omega_0 $ has at most two holes. Introducing the orthogonal projector $P^{(  m)}$ in $\mathscr H_n \otimes \mathscr H_N^-$ (acting trivially in $\mathscr H_n$) that projects onto the closed subspace
\begin{align}
\text{ran}P^{(  m)} = \Big\{ \Psi \in \mathscr H_n \otimes    \mathscr H_N^- \ : \sum_{k \in B_F} \sno a_k^* \Psi \sno^2  = m \sno \Psi \sno^2 \Big\}
\end{align}
(the subspace of all states with exactly $m$ holes), we obtain
\begin{align}
\Phi_3(t)  & =   \sum_{m=0}^2    i \int_0^t \D s\,  e^{i  \mathbb H s } P^{(  m )}  \mathbb V R \mathbb V  \xi(s) \otimes \Omega_0  =: \sum_{m=0}^2 \Phi_{3m}(t) \label{eq: Phi Interaction term 0ex} .
\end{align}

In the contribution $ \Phi_{32}(t)$ we need to expand a second time via integration by parts. Proceeding similarly as in \eqref{eq: Phi with resolvent}, one verifies
\begin{align}
\Phi_{32}(t) & =  \int_0^t \D s\, e^{i  \mathbb  H  s} e^{-i (T -E^0 +P_{\rm f}^2 + 1) s} \Big( \frac{d}{ds} e^{i( T - E^0+ P_{\rm f}^2 + 1 )s} \Big) R P^{  (2)} \mathbb  V R  \mathbb V \xi(s) \otimes \Omega_0 \nonumber \\[0mm]
&  =: \sum_{i=0}^3 \Phi_{32;i}(t) \label{eq: Phi Interaction term 2ex} 
\end{align}
with \allowdisplaybreaks
\begin{subequations}
\begin{align}
\Phi_{32;0}(t) & =  \Big[  e^{i  \mathbb H s}  R P^{  (j)}  \mathbb V  R \mathbb V  \xi(s) \otimes \Omega_0 \Big]_0^t  ,\label{eq: Phi boundary term 2} \\[1.5mm]
\Phi_{32;1}(t) & = - i \int_0^t \D s \, e^{i  \mathbb Hs} R P^{  (j)}  \mathbb V  R \mathbb V \big( -1 + h^0_n - h_n  \big) \xi(s) \otimes \Omega_0 ,  \label{eq: Phi boundary term 2b}  \\[1mm]
\Phi_{32;2}(t) & = - i \int_0^t  \D s\, e^{i   \mathbb H  s}   R P^{  (j)} \big\{ [ h^0_n , \mathbb V R \mathbb V ]  - P_{\rm f}^2  \mathbb  V R \mathbb V \big\}   \xi(s) \otimes \Omega_0  \label{eq: Phi h term 2}, \\[0.5mm]
\Phi_{32;3}(t) & = - i \int_0^t  \D s\, e^{i  \mathbb H  s}   \mathbb V  R P^{  (j)}  \mathbb V 	 R \mathbb  V  \xi(s) \otimes \Omega_0  . \label{eq: Phi Interaction term 2}
\end{align}
\end{subequations}
In the last line, we can decompose again in terms of the number of holes, i.e.
\begin{align}
\Phi_{32;3}(t) =  \sum_{m= 1}^{3} \Phi_{32;3m}(t),\quad \Phi_{32;3m}(t) = - i \int_0^t  \D s\, e^{i  \mathbb H  s}  P^{(m)} \mathbb V  R P^{  (2) }  \mathbb V 	 R \mathbb  V  \xi(s) \otimes \Omega_0 ,
\end{align}
since the state $\mathbb V  R P^{  (2 )} \mathbb V 	 R \mathbb  V  \xi(s) \otimes \Omega_0$ contains $m \in \{ 1,2,3\}$ holes.

Collecting everything we obtain the decomposition
\begin{align}
\Phi(t) & = \sum_{i=0}^2 \big( \Phi_i(t) +  \Phi_{32;i}(t) \big)  + \Phi_{30}(t) + \Phi_{31}(t) +  \sum_{m=1}^3\Phi_{32;3m}(t) .
\end{align}
Before we proceed, let us lay out the motivation behind this expansion. When estimating the separate contributions, the idea is that every $R$ should give a factor $k_F^{-1}$ whereas every $\mathbb V$ leads to a factor $\lambda k_F^{(d-1)/2}$ (modulo some log factors). This explains for instance why the norm of $\Phi_0(t)$ can be bounded by a constant times $\lambda k_F^{(d-2)/2}$ while the norm of $\Phi_{32;33}(t)$ can be bounded by a constant times $\vert t \vert \, \vert \lambda \vert^3 k_F^{(3d-7)/2}$. Even though this simple rule is the correct intuition, it also oversimplifies the situation somewhat as it is not applicable to each term in the expansion. (It does not apply whenever a $\mathbb V$ does not change the number of holes in the state it acts on, which happens for instance in $\Phi_{31}(t)$.)\medskip 

\noindent \textbf{Notation}. For the second part of the proof it is helpful to introduce further notation.\medskip

\noindent $\bullet$ For $i,j,u \in \{1,...,n\}$ we set
\begin{align}
K_{lk}^{(i)} = e^{i(k-l)y_i} ,\quad  K_{nm,lk}^{(i,j )} = K_{nm}^{(i)}  K_{lk}^{(j)} ,\quad 
K_{sr,nm,lk}^{(i,j,u)} = K_{sr}^{(i)}  K_{nm}^{(j)}  K_{lk}^{(u)} .
\end{align}
$\bullet$ To simply the notation, we write from now on $\widehat v_L(k) = \widehat v(k)$. Later on we shall also use the abbreviations
\begin{align}
\widehat v_{lk} = \widehat v(l-k) , \quad \varepsilon_{lk} = l^2-k^2 + (l-k)^2 .
\end{align}
\noindent $\bullet$ Moreover we write
\begin{align}
g(t, k_F ,L) \lesssim f(t, k_F ,L)
\end{align}
to indicate that there is a constant $C>0$ independent of the parameters $t$, $k_F$ and $L$ such that $g(t, k_F  ,L) \le C f(t, k_F ,L)$ for all $t\in \mathbb R $, $k_F \ge 2$ and $L> 0$. The constant $C$ is allowed to depend on the fixed model parameters $d$, $n$, $w$, $R$ and on the initial state $\xi_0$.

\paragraph{Estimates for the different contributions in $\Phi(t)+\phi(t)$}{\textcolor{white}{a}}\medskip

\noindent \textbf{Term $\Phi_0(t)$}. We use this first term to warm up with some simple computations. Since $a_l^* a_k \Omega_0$ is a simultaneous eigenstate of $T$ and $P_{\rm f}^2$ with eigenvalues $E_0+l^2-k^2$ and $(l-k)^2$, respectively, we have
\begin{align}
(T-E^0)a_l^* a_k \Omega_0 = (l^2-k^2)a_l^* a_k \Omega_0 , \quad P_{\rm f}^2  a_l^* a_k \Omega_0  = (l-k)^2a_l^* a_k \Omega_0,
\end{align}
and thus also
\begin{align}
R  a^*_l a_k \Omega_0 = (l^2-k^2 + (l-k)^2+ 1 )^{-1} a^*_l a_k \Omega_0.
\end{align} 
Applying the Wick rule \eqref{eq: Wick formula} as well as $\sno K_{lk}^{(i)}\xi(s) \sno_{\mathscr H _n} = \sno \xi(s) \sno_{\mathscr H _n} =  1$, one easily verifies
\begin{align}
& \sno  R \mathbb V^{(i)} \xi(s) \otimes \Omega_0 \sno^2  \\
& = L^{-2d}\sum_{(l,k)\in T_F} \sum_{(l',k')\in T_F}  \frac{\lambda^2  \overline{ \widehat v(l-k) } \widehat v(l'-k') \lsp K_{lk}^{(i)} \xi (s) , K_{l' k'}^{(i)} \xi(s) \rsp_{\mathscr H_n} \, \lsp a_l^* a_k \Omega_0, a_{l'}^* a_{k'} \Omega_0 \rsp_{\mathscr H_N^-}}  { (l^2 - k^2 +(l-k)^2+ 1 ) ({ l'}^2 - {k'}^2 +(l'-k')^2+ 1 ) }  \notag \\
& \le  L^{-2d}\sum_{(l,k)\in T_F} \frac{\lambda^2 \vert \widehat v(l-k)\vert^2 }{(l^2 - k^2 +(l-k)^2+ 1 )^2 } \quad \quad (\forall\, i = 1,...,n).
\end{align}

To bound the large volume limit of the remaining expression, we use \eqref{eq: sum of transitions 4} from Lemma \ref{lem: a priori bounds}. This leads to
\begin{align}
\limsup_{ L\to \infty  } \sno \Phi_0(t) \sno  \lesssim \vert  \lambda  \vert k_F^{(d-3)/2}   ( \ln k_F )^{1/2},
\end{align}
which provides the first error term in \eqref{def: GAMMA(d)}.\medskip

\noindent \textbf{Term $\Phi_1(t)$}. Following similar steps as in the above computation, one shows that
\begin{align}
& \sno  R \mathbb V^{(i)}  (-1  + h_n^0 - h_n )   \xi(s) \otimes \Omega_0 \sno^2 \nonumber \\[2mm]
& \hspace{2cm} \lesssim L^{-2d} \sum_{(l,k)\in T_F} \frac{\lambda^2 \vert \widehat v(l-k) \vert^2 }{( l^2-k^2+(l-k)^2 + 1)^2 } \big( 1 + \sno (h_n^0 -h_n )\xi(s) \sno^2_{\mathscr H_n} \big) . \label{eq: Phi boundary term 2}
\end{align}
To bound the remaining norm, use $h_n^0 - h_n = \sum_{i<j}^n \lambda^2 W_{k_F}(y_i-y_j ) + n  \lambda^2 W_{k_F}(0)$ and $\vert W_{k_F}( r )  \vert \le  W_{k_F}( 0 )$. Hence we can apply \eqref{eq: sum of transitions 2} and \eqref{eq: sum of transitions 4} to find the bound
\begin{align}
\limsup_{L\to \infty} \sno \Phi_1(t) \sno  \lesssim  \vert t \vert (1 + \lambda^2 k_F^{(d-2)} ) \vert \lambda \vert k_F^{(d-3)/2} (\ln k_F)^{1/2}.
\end{align}
\textbf{Term $\Phi_2(t)$}. Here we need to evaluate
\begin{align}
P_{\rm f}^2 \mathbb V^{(i)}   \xi(s) \otimes \Omega_0   &  = L^{-d} \sum_{(l,k)\in T_F} \widehat v(l-k) (l-k)^2 K_{lk}^{(i)} \xi(s) \otimes  a^*_l a_k\Omega_0
\end{align}
and
\begin{align}
 [ h^0_n , \mathbb V^{(i)} ] \xi(s) \otimes \Omega_0 &  =  L^{-d} \sum_{(l,k)\in T_F} \widehat v(l-k)  ( l-k )^2 K_{lk}^{(i)}   \xi(s)  a^*_l a_k\Omega_0\nonumber \\
& \quad +   L^{-d} \sum_{(l,k)\in T_F} \widehat v(l-k)  K_{lk}^{(i)}   (k-l) \cdot (-2  i\nabla_{y_i})  \xi(s) \otimes a^*_l a_k\Omega_0.
 \end{align}
Taking the difference, the terms proportional to $(l-k)^2$ cancel out, which is the reason for including $P_{\rm f}^2$ in $R$ (see the remark in Footnote \ref{footnote: pf squared}). Thus we obtain
\begin{align}
& R \big\{ [ h^0_n , \mathbb V^{(i)} ] - P_{\rm f}^2\mathbb V^{(i)} \big\}  \xi(s) \otimes \Omega_0  \nonumber \\[2mm]
& \quad \quad  \quad =  L^{-d} \sum_{(l,k)\in T_F} \widehat v(l-k)  K_{lk}^{(i)}   (k-l) \cdot (-2i\nabla_{y_i})   \xi(s) \otimes  R a^*_l a_k\Omega_0,
\end{align}
the norm of which we can estimate by
\begin{align}
& \sno  R \big\{ [ h^0_n , \mathbb V^{(i)} ] - P_{\rm f}^2\mathbb V^{(i)} \big\}  \xi(s) \otimes \Omega_0 \sno ^2 \nonumber \\[2mm]
& \quad \quad \quad \lesssim \bigg( L^{-2d} \sum_{(l,k)\in T_F}  \frac{ \lambda^2 \vert \widehat v(l-k)\vert^2 (l-k)^2}{(l^2-k^2+(l-k)^2+1)^2} \bigg) \sno \nabla_{y_i }  \xi(s)\sno^2_{\mathscr H_n}.\label{eq: term b}
\end{align}

To bound the norm involving the gradient we use the assumption that $w^2 \le c (1-\Delta)$ for some $ c\in [0,1) $. This implies
\begin{align}
\sum_{i=1}^n \lsp \xi(s),  (- \Delta_{y_i}) \xi(s) \rsp_{\mathscr H_n} \lesssim  1 + \lsp \xi(s), h_n^0 \xi(s) \rsp_{\mathscr H_n},
\end{align}
and since $\vert W_{k_F}(r)\vert \le W_{k_F}(0)\lesssim k_F^{(d-2)}$, we can proceed by 
\begin{align}\label{eq: bound for kinetic energy}
\lsp \xi(s), h_n^0 \xi(s) \rsp & \lesssim  \lambda^2 k_F^{(d-2)} +    \lsp \xi(0), h_n^0 \xi(0) \rsp_{\mathscr H_n} \nonumber \\
&   \lesssim  \lambda^2 k_F^{(d-2)} + \sum_{i=1}^n \lsp \xi(0),  (- \Delta_{y_i}) \xi(0) \rsp_{\mathscr H_n} \lesssim  \lambda^2 k_F^{(d-2)} +1,
\end{align}
where the last step follows from Assumption \eqref{eq: condition on the kinetic energy}. 

In combination with \eqref{eq: sum of transitions 4} we can now take the large-volume limit in \eqref{eq: term b} to find
\begin{align} 
\limsup_{ L\to \infty  } \sno \Phi_2(t) \sno \lesssim \vert t \vert (1+\lambda^2 k_F^{(d-2)})  \vert \lambda \vert k_F^{(d-3)/2}  (\ln k_F)^{1/2}  .
\end{align}
\textbf{Term $\phi(t) + \Phi_{30}(t)$}. The contribution $\Phi_{30}(t)$ is the one that determines the effective Hamiltonian $h_n$. To see this, use
\begin{align}
\lsp \Omega_0, \mathbb V^{(i)}  R \mathbb V^{(j)} \Omega_0 \rsp_{\mathscr H_N^- }  = L^{-2d} \sum_{(l,k)\in T_F} \frac{\vert \widehat v(l-k) \vert^2}{l^2-k^2 + (l-k)^2 + 1 } e^{i(l-k)(y_j-y_i)},
\end{align}
and $P^{(0)}  = \id \otimes \vert \Omega_0 \rangle \langle \Omega_0 \vert$, in order to compute 
\begin{subequations}
\begin{align}
& \Phi_{30}(t)  =  \sum_{i,j = 1}^n i \int_0^t \D s\,  e^{i  \mathbb H   s} P^{  (0)} \mathbb V^{(i)}  R \mathbb V^{(j)} \xi(s) \otimes \Omega_0  \nonumber\\
&   = i  \sum_{i=1}^n  \int_0^t \D s\, e^{i  \mathbb  H  s} \Bigg(  L^{-2d}\sum_{(l,k)\in T_F}  \frac{ \lambda^2 \vert \widehat v(l-k)\vert^2}{ l^2-k^2 + (l-k)^2  +1 }  \Bigg)  \xi(s) \otimes \Omega_0 \label{eq: Phi Interaction term 0ex line 1}\\[0mm]
&  + i\sum_{i < j  }^n   \int_0^t \D s\,  e^{i  \mathbb  H  s} \Bigg( L^{-2d}\sum_{(l,k)\in T_F} \frac{ \lambda^2 \vert \widehat v(l-k)\vert^2}{ l^2-k^2 +  (l-k)^2 + 1 }  \cos((k-l) \cdot (y_i-y_j))  \Bigg) \xi(s) \otimes \Omega_0 .\label{eq: Phi Interaction term 0ex line 2}
\end{align}
\end{subequations}
In the large-volume limit, the expressions in parenthesis converge to $\lambda^2 W_{k_F }(0) $ and $\lambda^2 W_{k_F }(\vert y_i-y_j\vert )$, respectively (since the Riemann sums converge to the corresponding integrals). Because $\widehat v_\infty(k)$ is rotational invariant we can then replace the argument in the cosine by $(l-k) \cdot \hat a \vert y_i-y_j \vert$ for any unit vector $\hat a \in \mathbb R^d$. Since
\begin{align}
\phi(t) =  i  \int_0^t \D s\, e^{i  \mathbb  H  s} \bigg( -  \sum_{i<j}^n \lambda^2 W_{k_F}(\vert y_i-y_j\vert ) - n \lambda^2  W_{k_F}(0) \bigg) \xi(s) \otimes \Omega_0 ,
\end{align}
we get a complete cancellation between $\Phi_{30}(t)$ and $\phi(t)$, that is
\begin{align}
\lim_{ L\to \infty  } \sno \phi(t) + \Phi_{30}(t)  \sno =0 .
\end{align}

We emphasize that this is crucial since the norm of $\Phi_{30}(t)$ is of order $\lambda^2 k_F^{(d-2)}$ which is of order one if we choose $\lambda^{2} = k_F^{(2-d)}$.\medskip

\noindent \textbf{Term $\Phi_{31}(t)$}. Abbreviating $\varepsilon_{lk}=l^2-k^2+(l-k)^2 $ and $\widehat v_{lk} = \widehat v(l-k)$, we compute
\begin{subequations}
\begin{align}
  P^{( 1)} \mathbb V^{(i)} R \mathbb V^{(j)}   \xi(s) \otimes \Omega_0   & =  L^{-2d} \sum_{m\in B_F} \sum_{(l,k) \in T_F} \frac{\lambda^2 \widehat v_{km} \widehat v_{lk} }{\varepsilon_{lk} + 1} K_{km,lk}^{(i,j)} \xi(s)\otimes a_m a_l^* \Omega_0 \label{eq: Phi Interaction term 1ex line 1} \\[1mm]
&  +  L^{-2d} \sum_{n\in (B_F)^c} \sum_{(l,k) \in T_F} \frac{\lambda ^2 \widehat v_{nl} \widehat v_{lk} }{\varepsilon_{lk} +1 } K_{nl,lk}^{(i,j)}\xi(s)\otimes a_n^* a_k \Omega_0,\label{eq: Phi Interaction term 1ex line 2}
\end{align}
\end{subequations}
where we utilized the identity
\begin{align}
P^{(1)} a^*_n a_m a_l^* a_k \Omega_0  = \delta_{kn}\chi_{B_F}(m) \, a_m a_l^* \Omega_0 + \delta_{ml} \chi_{B_F^c}(n) \, a_n^* a_k \Omega_0 \quad \forall  \, (l,k) \in B_F^c \times B_F .
\end{align}
Using $ \sno K_{km,lk}^{(i,j)} \xi(s) \sno_{\mathscr H_n} = 1$ we proceed in the first line with
\begin{align}
\sno \eqref{eq: Phi Interaction term 1ex line 1} \sno^2 &  \le L^{-4d} \sum_{m,m'\in B_F} \sum_{\substack{ (l,k) \in T_F } } \sum_{\substack{  (l',k') \in T_F } } \frac{\lambda^4 \vert \widehat v_{km} \vert \, \vert \widehat v_{ k'm'} \vert \, \vert\widehat v_{lk} \vert \vert\widehat v_{l'k'} \vert }{ (\varepsilon_{lk} +1 )(\varepsilon_{l'k'}+1)} \Big\vert \lsp a_{m'} a_{l'}^* \Omega_0 , a_m a_l^* \Omega_0 \rsp_{\mathscr H_N^-} \Big\vert\nonumber \\
& = \lambda^4 \Bigg( L^{-2d} \sum_{(l,m)\in T_F} \bigg( L^{-d} \sum_{k\in B_F} \frac{ \vert\widehat v(l-k) \vert\, \vert \widehat  v(k-m) \vert  }{ l^2-k^2 + (l-k)^2 +  1 } \bigg)^2 \Bigg) ,
\end{align}
and similarly in the second line with 
\begin{align}
 \sno \eqref{eq: Phi Interaction term 1ex line 2} \sno^2 & \le \lambda^4 \Bigg( L^{-2d} \sum_{(n,k )\in T_F} \bigg( L^{-d} \sum_{l\in B_F^c} \frac{\vert\widehat v(l-k)  \vert \, \vert \widehat v(n-l) \vert  }{l^2-k^2+ (l-k)^2 + 1 } \bigg)^2 \Bigg)  . 
\end{align}

The two remaining expressions are estimated in Lemma \ref{lem: a priori bounds 2}, \eqref{eq: lem 2 bound 5} and \eqref{eq: lem 2 bound 6}. This implies
\begin{align}
\limsup_{ L\to \infty  } \sno \Phi_{21}(t) \sno \lesssim  \vert t\vert   \lambda^2  \sqrt{ \gamma(d,k_F) }
\end{align}
with $\gamma(d,k_F)$ defined in \eqref{def: Gamma(d)}.\medskip

\noindent \textbf{Term $\Phi_{32;0}(t)$}. Here we have
\begin{align}
& P^{  (2)} R \mathbb V^{(i )}  R  \mathbb V^{(j )} \xi(s) \otimes \Omega_0 \nonumber \\[2mm]
& \qquad = L^{-2d}\sum_{(n,m) \in T_F} \sum_{(l,k) \in T_F} \frac{\lambda^2 \widehat v_{nm} \widehat v_{lk} }{( \varepsilon_{nm} + \varepsilon_{lk}  + 1  ) ( \varepsilon_{lk}+ 1     ) } K_{nm,lk}^{(i,j)} \xi(s) \otimes a_n^* a_m a^*_l a_k \Omega_0,
\end{align}
which follows from
\begin{align}
P^{(2)}  a_n^* a_m a_l^* a_k \Omega_0  = \chi_{T_F}( n,m)  \,  a_n^* a_m a_l^* a_k \Omega_0 \quad \forall (l,k) \in B_F^c\times B_F.
\end{align}
Using the basic inequality
\begin{align}
& \Big\vert \frac{  \widehat v_{n'm'}  \widehat v_{l'k'} }{( \varepsilon _{n'm'}+ \varepsilon_{l'k'}+ 1   )(\varepsilon_{l'k'}+ 1  )} \,  \frac{  \widehat v_{nm}  \widehat v_{lk} }{( \varepsilon _{nm}+ \varepsilon_{lk}+ 1   ) (\varepsilon_{lk}+ 1  )} \Big\vert \nonumber \\[1.5mm]
& \hspace{3.5cm} \le \frac{ \vert  \widehat v_{n'm'}  \widehat v_{'l'k}\vert^2}{2 ( \varepsilon _{n'm'} + 1   )^2 (\varepsilon_{l'k'}+ 1  )^2}  + \frac{ \vert   \widehat v_{nm}  \widehat v_{lk}\vert^2}{ 2 ( \varepsilon _{nm} + 1   )^2 (\varepsilon_{lk}+ 1  )^2} ,
\end{align}
we can estimate the norm by 
\begin{align}\label{eq: estimate P2}
\sno P^{  (2)} R \mathbb  V^{(i)}  R \mathbb  V^{(j )} \xi(s) \otimes \Omega_0 \sno^2&  \le L^{-4d} \sum_{\substack{ (n,m) \in T_F \\ (n',m') \in T_F }} \sum_{\substack { (l,k) \in T_F \\ (l',k') \in T_F } } \frac{\lambda^4 \vert \widehat v_{nm} \widehat v_{lk} \vert }{( \varepsilon_{nm} + \varepsilon_{lk} + 1      )^2  ( \varepsilon_{lk} + 1  )^2  }  \nonumber \\[-2mm] 
& \hspace{4cm} \times \Big\vert \lsp  a^*_{n'} a_{m'} a^*_{l'} a_{k'} \Omega_0 ,  a^*_n a_m a^*_l a_k \Omega_0 \rsp_{\mathscr H_N^-} \Big\vert \notag \\
&  \le 4 \Bigg( L^{-2d} \sum_{(l,k) \in T_F}  \frac{\lambda^2\vert \widehat v_{lk} \vert^2 }{ (l^2-k^2 + 1  )^2} \Bigg)^2.
\end{align}
In the last step we used that the scalar product provides four different possibilities to cancel the primed summation.

With the aid of Lemma \ref{lem: a priori bounds} we get
\begin{align}\label{eq: not sufficient bound for d=1 B}
\limsup_{ L \to \infty } \sno \Phi_{32;0}(t) \sno \lesssim  \lambda^2   k_F^{(d-3)} \ln k_F .
\end{align}
\textbf{Term $\Phi_{32;1}(t)$}. Proceeding similarly as in the previous computation, we obtain
\begin{align}
& \sno R P^{  (2)}  \mathbb V^{(i)}  R \mathbb V^{(j)} \big( -1 + h_0 - h  \big) \xi(s) \otimes \Omega_0 \sno^2 \nonumber \\[1.5mm]
& \quad \quad \quad \lesssim  L^{-4d} \sum_{(n,m) \in T_F} \sum_{(l,k) \in T_F} \Big\vert\frac{\lambda^2 \widehat v_{nm} \widehat v_{lk} }{( \varepsilon_{nm} + \varepsilon_{lk} + 1      ) ( \varepsilon_{lk} + 1  ) } \Big\vert^2  (1 + \sno (h_n-h^0_n) \xi_0 \sno^2_{\mathscr H_n}) .
\end{align}
From here we can follow analogous steps as for $\Phi_1(t)$, in order to find
\begin{align}
\limsup_{L\to \infty} \sno \Phi_{32;1}(t) \sno \lesssim \vert t \vert (1 + \lambda^2 k_F^{(d-2)} ) \lambda^2 k_F^{(d-3)}\ln k_F.
\end{align}
\noindent\textbf{Term $\Phi_{32;2}(t)$}.
Similarly as in $\Phi_2(t)$ , here it is important that certain contributions cancel each other. To see this, we compute
\begin{align}
 P^{(  2)} [ h^0 , \mathbb V R \mathbb V ]\Omega_0  & = \sum_{i,j =1}^n  \mathbb V^{(i)} R \big[ h^0 , \mathbb   V^{(j)} \big]\Omega_0 +  R \big[ h^0 , \mathbb  V^{(i)}  \big] \mathbb  V^{(j)} \Omega_0 \nonumber \\[1mm]
 &  = \sum_{i,j =1}^2 L^{-2d}\sum_{(l,k)\in T_F}\sum_{(n,m)\in T_F} \widehat v_{nm} \widehat v_{lk}   K_{nm,lk}^{(i,j)} ( l-k + n - m )^2  a^*_n a_m R a^*_l a_k \Omega_0 \nonumber \\[1mm]
  & \quad + \sum_{i,j=1}^n L^{-2d}\sum_{(l,k)\in T_F}\sum_{(n,m)\in T_F} \widehat v_{nm}  \widehat v_{lk}   G_{nm,lk}^{(i,j)} a^*_n a_m R a^*_l a_k \Omega_0
\end{align}
where 
\begin{align}
 G_{nm,lk}^{(i,j)} =   K_{nm,lk}^{(i,j) }  \big( 2 (n-m) \cdot (- i \nabla_{y_i})  + 2 (l-k) \cdot (-i\nabla_{y_j}) \big) .
\end{align}
If we subtract
\begin{align}
&  P_{\rm f}^2 P^{( 2)} \mathbb V R \mathbb V \Omega_0   =\sum_{i,j =1}^n L^{-2d}\sum_{(l,k)\in T_F}\sum_{(n,m)\in T_F} \widehat v_{nm} \widehat v_{lk} K_{nm,lk}^{(i,j)}  (l-k+n-m)^2 a^*_n a_m R a^*_l a_k \Omega_0,
\end{align}
the terms proportional to $(l-k+n-m)^2$ cancel each other. The difference can thus be bounded by
\begin{align}
& \sno   R  \big\{ P^{  (2)} \big[ h^0 ,  \mathbb V   R \mathbb V \big]  - P_{\rm f}^2 P^{  (2)} \mathbb V R \mathbb V \big\} \xi(s) \otimes \Omega_0 \sno^2 \nonumber \\[2mm]
& \ \lesssim L^{-4d} \sum_{(n,m) \in T_F} \sum_{(l,k) \in T_F}  \frac{ \lambda^4\vert \widehat v_{nm} \widehat v_{lk}  \vert^2 }{( \varepsilon_{nm} + \varepsilon_{lk}+  1    )^2 ( \varepsilon_{lk}+  1  )^2 }  \big( (  n-m )^2  + ( l- k ) ^2 \big)  \sum_{i=1}^n \sno   \nabla_{y_i}  \xi(s) \sno^2_{\mathscr H_n} \nonumber \\
& \ \lesssim \lambda^4 \bigg( L^{-2d} \sum_{(l,k) \in T_F} \frac{ \vert \widehat v_{l k } \vert^2 (  l-k ) ^2 }{ ( l^2-k^2+(l-k)^2+1 ) ^2}\bigg) \bigg( L^{-2d} \sum_{(l,k) \in T_F} \frac{  \vert \widehat v_{lk} \vert^2 }{ ( l^2-k^2 +1 ) ^2}\bigg)(1+\lambda^2 k_F^{(2-d)})
\end{align}
where we employed \eqref{eq: bound for kinetic energy} in the last step. Using Lemma \ref{lem: a priori bounds} we get
\begin{align}
\limsup_{ L\to \infty  } \sno \Phi_{32;1}(t) \sno  \lesssim   \vert t\vert (1+\lambda^2 k_F^{(2-d)})  \lambda^2 k_F^{(d-3)}  (\ln k_F )^{3/2}.
\end{align}
\textbf{Term $\Phi_{32;33}(t)$}. We have
\begin{align}
& P^{ (3)}  R  \mathbb V^{(i)}  P^{(2)} \mathbb  V^{(j)}	 R \mathbb  V^{(k)} \xi(s) \otimes \Omega_0 \nonumber \\[3mm]
& = L^{-3d}\sum_{(s,r ) \in T_F} \sum_{(n,m) \in T_F} \sum_{(l,k) \in T_F} \frac{\lambda^3 \widehat v_{sr}  \widehat v_{nm}  \widehat v_{lk}}{( \varepsilon _{nm}+ \varepsilon_{lk}+ 1 ) (\varepsilon_{lk} + 1 )} K_{sr,nm,lk}^{(i,j,k)} \xi(s) \otimes a_s^*a_r a^*_n a_m a^*_l a_k \Omega_0 
\end{align}
which is a direct consequence of
\begin{align}
P^{(3)} a^*_s a_r a_n^* a_m a_l^* a_k \Omega_0 = \chi_{T_F}( s,r )  \chi_{T_F}( n,m )  \, a_n^* a_m a_l^* a_k \Omega_0  \quad \forall (l,k) \in B_F^c\times B_F.
\end{align}
Using
\begin{align}
& \Big\vert \frac{  \widehat v_{s'r' }  \widehat v_{n'm'}  \widehat v_{l'k'} }{( \varepsilon _{n'm'}+ \varepsilon_{l'k'}+ 1   )(\varepsilon_{l'k'}+ 1  )} \frac{   \widehat v_{sr }  \widehat v_{nm}  \widehat v_{lk} }{( \varepsilon _{nm}+ \varepsilon_{lk}+ 1   ) (\varepsilon_{lk}+ 1  )} \Big\vert \nonumber \\[1.5mm]
& \hspace{3.5cm} \le \frac{ \vert \widehat v_{s'r'}  \widehat v_{n'm'}  \widehat v_{'l'k}\vert^2}{2 ( \varepsilon _{n'm'} + 1   )^2 (\varepsilon_{l'k'}+ 1  )^2}  + \frac{ \vert \widehat v_{sr}  \widehat v_{nm}  \widehat v_{lk}\vert^2}{ 2 ( \varepsilon _{nm} + 1   )^2 (\varepsilon_{lk}+ 1  )^2} ,
\end{align}
we proceed by\allowdisplaybreaks
\begin{align}
& \sno P^{\rm (3)}  R  \mathbb V^{(i)}   P^{\rm (2)} \mathbb V^{(j)}	 R \mathbb  V^{(u)} \xi(s) \otimes \Omega_0 \sno ^2\nonumber \\[3mm]
& \hspace{-2mm} \le L^{-6d} \sum_{\substack{ (s,r) \in T_F \\  (s',r') \in T_F} } \sum_{\substack{ (n,m) \in T_F \\  (n',m') \in T_F} } \sum_{\substack{ (l,k) \in T_F \\  (l',k') \in T_F} } \frac{\lambda^6 \vert \widehat v_{sr }  \widehat v_{nm}  \widehat v_{lk}\vert^2}{(  \varepsilon_{nm}+ 1   )^2 (\varepsilon_{lk}+ 1  )^2} \sno  K_{sr ,nm,lk}^{(i,j,u)} \xi(s) \sno^2_{\mathscr H_n} \nonumber \\[-2mm] 
& \hspace{6.5cm} \times \Big\vert \lsp a_{s'}^*a_{r'} a^*_{n'} a_{m'} a^*_{l'} a_{k'} \Omega_0 , a_s^*a_r a^*_n a_m a^*_l a_k \Omega_0 \rsp_{\mathscr H_N^-} \Big\vert .\label{eq: P3 term}
\end{align}
In the last expression we employ $\sno  K_{sr ,nm,lk}^{(i,j,u)} \xi(s) \sno_{\mathscr H_n} = 1$ and use the fact that the scalar product provides $36$ different combinations to cancel the primed summations. This gives
\begin{align}
\eqref{eq: P3 term}\lesssim \lambda^6 \Bigg( L^{-2d} \sum_{(s,r) \in T_F} \vert \widehat v_{sr} \vert^2 \Bigg) \Bigg( L^{-2d}\sum_{(n,m) \in T_F}   \frac{ \vert \widehat v_{nm}  \vert^2}{( \varepsilon_{nm} + 1    )^2 }\Bigg) \Bigg(  L^{-2d} \sum_{(l,k) \in T_F}     \frac{ \vert \widehat v_{lk} \vert^2 }{( \varepsilon_{lk}+ 1 )^2 }\Bigg).
\end{align}
By Lemma \ref{lem: a priori bounds} we obtain
\begin{align}
\limsup_{ L\to \infty  } \sno \Phi_{22;23}(t) \sno  \lesssim \vert t \vert \, \vert \lambda \vert^3 k_F ^{ ( 3d-7 ) / 2 }  \ln k_F .
\end{align}
\noindent \textbf{Term $\Phi_{32;31}(t) $}. Next we consider the contributions with one hole. Here one verifies
\begin{subequations}
\begin{align}
&\hspace{-1cm} P^{  (1)}  \mathbb V^{(i)}  P^{  (2)} R \mathbb V^{(j)} 	 R  \mathbb V^{(u)}   \xi(s) \otimes \Omega_0 \nonumber \\[3.5mm]
&  \hspace{-2.5mm} = 2 L^{-3d}  \sum_{(n,m) \in T_F} \sum_{(l,k) \in T_F} \frac{\lambda^3 \vert \widehat v_{mn} \vert^2 \widehat v_{lk}}{( \varepsilon _{nm}+ \varepsilon_{lk}+ 1 ) (\varepsilon_{lk} + 1 ) } K_{mn,nm,lk}^{(i,j,u)}  \xi(s) \otimes a^*_l a_k \Omega_0  \label{eq: 3 pi one excitation line 1} \\
&  \hspace{-2.5mm} \quad + 2 L^{-3d} \sum_{(n,m) \in T_F} \sum_{(l,k) \in T_F} \frac{\lambda^3 \widehat v_{nm}  \vert \widehat v_{lk} \vert^2}{( \varepsilon _{nm}+ \varepsilon_{lk}+ 1 ) (\varepsilon_{lk}+ 1 )}  K_{kl,nm,lk}^{(i,j,u)} \xi(s) \otimes   a_n^*   a_m \Omega_0 \label{eq: 3 pi one excitation line 2} \\
&  \hspace{-2.5mm} \quad + 2 L^{-3d} \sum_{(n,m) \in T_F} \sum_{(l,k) \in T_F} \frac{ \lambda^3 \widehat v_{kn}  \widehat v_{nm}  \widehat v_{lk}}{( \varepsilon _{nm}+ \varepsilon_{lk}+ 1 ) (\varepsilon_{lk}+ 1 ) }  K_{kn,nm,lk}^{(i,j,u) } \xi(s) \otimes   a_m a^*_l  \Omega_0  \label{eq: 3 pi one excitation line 3}   \\
&  \hspace{-2.5mm} \quad - 2 L^{-3d}  \sum_{(n,m) \in T_F} \sum_{(l,k) \in T_F} \frac{\lambda^3 \widehat v_{ml}  \widehat v_{nm}  \widehat v_{lk} }{( \varepsilon _{nm}+ \varepsilon_{lk}+ 1  ) (\varepsilon_{lk}+ 1 )}  K_{ml,nm,lk}^{(i,j,u)} \xi(s) \otimes a_n^* a_k \Omega_0  
\label{eq: 3 pi one excitation line 4}  
\end{align}
\end{subequations}
since
\begin{align}
P^{(1)} a^*_s a_r a_n^* a_m a_l^* a_k \Omega_0 & = \delta_{sm} \delta_{rn} a^*_l a_k \Omega_0 +  \delta_{sk} \delta_{rl} a_n^* a_m \Omega_0\notag \\
&  \quad  + \delta_{sk} \delta _{rn} a_m a_l^*\Omega_0 + \delta_{sm} \delta_{rl} a_n^* a_k \Omega_0 \quad  \forall l,n\in B_F^c, \ k,m \in B_F.
\end{align}

%\begin{align}
%& \hspace{-2.5mm} + L^{-6d}  \sum_{(n,m) \in S_F} \sum_{(l,k) \in S_F} \frac{\widehat w(m-n)  \widehat w(n-m)  \widehat w(l-k)}{(n^2-m^2+l^2-k^2 )(l^2-k^2)} \big( k_{mn}k_{nm}k_{lk}\xi(s) \otimes   a^*_l a_k \Omega \big) \\[1mm]
%& \hspace{-2.5mm} - L^{-6d}  \sum_{(n,m) \in S_F} \sum_{(l,k) \in S_F} \frac{\widehat w(m-l)  \widehat w(n-m)  \widehat w(l-k)}{(n^2-m^2+l^2-k^2 )(l^2-k^2)} \big( k_{ml}k_{nm}k_{lk}\xi(s) \otimes   a^*_n   a_k \Omega \big) \\[1mm]
%&  \hspace{-2.5mm} +  L^{-6d}  \sum_{(n,m) \in S_F} \sum_{(l,k) \in S_F} \frac{\widehat w(k-n)  \widehat w(n-m)  \widehat w(l-k)}{(n^2-m^2+l^2-k^2 )(l^2-k^2)} \big( k_{kn}k_{nm}k_{lk}\xi(s) \otimes   a_m a^*_l \Omega \big) \\[1mm]
%&  \hspace{-2.5mm} +  L^{-6d} \sum_{(n,m) \in S_F} \sum_{(l,k) \in S_F} \frac{\widehat w(k-l)  \widehat w(n-m)  \widehat w(l-k)}{(n^2-m^2+l^2-k^2 )(l^2-k^2)} \big( k_{kl}k_{nm}k_{lk}\xi(s) \otimes  a^*_n  a_m  \Omega \big) 
%\end{align}\\
The first line is estimated by
\begin{align}
\sno \eqref{eq: 3 pi one excitation line 1} \sno^2  \le  4 \lambda^6 \Bigg( L^{-2d}  \sum_{(n,m)\in T_F} \frac{\vert \widehat v(n-m)\vert^2}{m^2-n^2 + 1  } \Bigg)^2   \Bigg( L^{-2d} \sum_{(l,k) \in T_F} \frac{\vert \widehat v(l-k)\vert^2}{(l^2-k^2 + 1 )^2}  \Bigg),
\end{align}
and the same bound holds for $\sno  \eqref{eq: 3 pi one excitation line 2}\sno^2$. Similarly one finds
\begin{align}
\sno \eqref{eq: 3 pi one excitation line 3}\sno ^2 \le 4\lambda^6  \Bigg( L^{-2d} \sum_{(n',k') \in T_F} \vert \widehat v(n'-k')\vert^2\Bigg) \Bigg( L^{-2d}\sum_{(l,k)\in T_F} \frac{\vert \widehat v(l - k)\vert^2 }{(l^2-k^2+ 1 )^2}\Bigg)^2
\end{align}
which holds as well for $\sno \eqref{eq: 3 pi one excitation line 4} \sno ^2$. In total we get
\begin{align}
\limsup_{ L\to \infty  } \sno \Phi_{32;31}(t) \sno  \lesssim  \vert t\vert \, \vert \lambda \vert^3 k_F^{ ( 3d - 7 ) / 2  } ( \ln k_F )^{1/2}.
\end{align}
\textbf{Term $\Phi_{32;32}(t)$}. Lastly we come to the contributions with two holes. A straightforward computation leads to\allowdisplaybreaks
\begin{subequations}
\begin{align}
& \hspace{-0.25cm} P^{  (2)}  \mathbb V^{(i)}  P^{  (2)}  R \mathbb V^{(j)} 	 R  \mathbb  V^{(u)} \xi(s) \otimes \Omega_0 \nonumber \\[1mm]
&  \hspace{-2.5mm} = L^{-3d}\sum_{ s\in B_F^c } \sum_{(n,m) \in T_F} \sum_{(l,k) \in T_F} \frac{\lambda^3 \widehat v_{rn}  \widehat v_{nm}  \widehat v_{lk}}{( \varepsilon _{nm}+ \varepsilon_{lk} + 1  ) (\varepsilon_{lk}+ 1  ) } K_{sn,nm,lk}^{(i,j,u)} \xi(s) \otimes a_s^* a_m a^*_l a_k \Omega_0  \label{eq: 3 pi two excitations line 1}\\ 
&  \hspace{-2.5mm} \quad + L^{-3d}\sum_{ s\in B_F^c } \sum_{(n,m) \in T_F} \sum_{(l,k) \in T_F} \frac{\lambda^3 \widehat v_{sl}  \widehat v_{nm}  \widehat v_{lk}}{( \varepsilon _{nm}+ \varepsilon_{lk}+ 1 ) (\varepsilon_{lk}+  1  )}  K_{rl,nm,lk}^{(i,j,u)} \xi(s) \otimes a_s^* a_n^* a_m a_k \Omega_0 \label{eq: 3 pi two excitations line 2} \\ 
&  \hspace{-2.5mm} \quad + L^{-3d}\sum_{ r \in B_F } \sum_{(n,m) \in T_F} \sum_{(l,k) \in T_F} \frac{\lambda^3 \widehat v_{mr}  \widehat v_{nm} \widehat v_{lk} }{( \varepsilon _{nm}+ \varepsilon_{lk}+ 1 ) (\varepsilon_{lk}+  1 )}  K_{mr,nm,lk}^{(i,j,u)}\xi(s) \otimes a_r a^*_n  a^*_l a_k \Omega_0  \label{eq: 3 pi two excitations line 3}\\
&  \hspace{-2.5mm} \quad +  L^{-3d}\sum_{ r\in B_F } \sum_{(n,m) \in T_F} \sum_{(l,k) \in T_F} \frac{\lambda^3 \widehat v_{kr}  \widehat v_{nm}  \widehat v_{lk}}{( \varepsilon _{nm}+ \varepsilon_{lk}+  1 ) (\varepsilon_{lk}+  1  )}  K_{ks,nm,lk}^{(i,j,u)}  \xi(s) \otimes a_r a^*_n  a_m a^*_l \Omega_0 .\label{eq: 3 pi two excitations line 4}
\end{align}
\end{subequations}
where we used
\begin{align}
P^{(2)} a^*_s a_r a_n^* a_m a_l^* a_k \Omega_0 & = \chi_{B_F^c}(s) \big( \delta_{rn}\, a^*_s a_m a_l^* a_k \Omega_0 + \delta_{rl}  a^*_s a_n^* a_m  a_k \Omega_0 \big)  \notag\\[1mm]
& \hspace{-2cm} + \chi_{B_F}(r) \big( \delta_{sm} \,  a_r a_n^* a^*_l a_k \Omega_0 +  \delta_{sk}\,  a_r a_n^* a_m a_l^* \Omega_0\big)  \quad \forall l,n\in B_F^c, \ k,m \in B_F.
\end{align}
We estimate
\begin{subequations}
\begin{align}
\sno \eqref{eq: 3 pi two excitations line 1}\sno ^2  & \le \bigg( L^{-2d}\sum_{(l,k)\in T_F} \frac{\lambda^2 \vert \widehat v_{lk}\vert^2}{( \varepsilon_{lk} + 1 )^2 } \bigg) \bigg( L^{-2d}\sum_{(s,m ) \in T_F} \bigg( L^{-d}\sum_{n \in B_F^c} \frac{ \lambda^2 \vert \widehat v_{nm} \vert\, \vert \widehat v_{ns}\vert }{\varepsilon_{nm} + 1 } \bigg)^2 \bigg) \\
& \quad  +  L^{-2d}\sum_{ s , l \in B_F^c} \bigg( L^{-2d} \sum_{(n,m) \in T_F} \frac{\lambda^3 \vert \widehat v_{lm}\vert \, \vert \widehat v_{nm} \vert \, \widehat v_{sn} \vert }{ ( \varepsilon_{lm} + 1 ) ( \varepsilon_{nm} + 1 ) } \bigg)^2\\
& \quad +  L^{-2d}\sum_{ k,m \in B_F} \bigg( L^{-2d} \sum_{ l, n \in B_F^c } \frac{\lambda^3 \vert \widehat v_{lk}\vert \, \vert \widehat v_{n m} \vert \, \widehat v_{ln} \vert }{ ( \varepsilon_{lk} + 1 ) ( \varepsilon_{nm} + 1 ) } \bigg)^2\\
& \quad + \bigg(  L^{-3d}\sum_{ (l,k) \in T_F } \sum_{n \in B_F}   \frac{\lambda^3 \vert \widehat v_{lk}\vert \, \vert \widehat v_{n k} \vert \, \widehat v_{ln} \vert }{ ( \varepsilon_{lk} + 1 ) ( \varepsilon_{n k} + 1 ) } \bigg)^2,
\end{align}
\end{subequations}
and in close analogy, one derives the same bound also for \eqref{eq: 3 pi two excitations line 2}-\eqref{eq: 3 pi two excitations line 4}.

With the aid of Lemma \ref{lem: a priori bounds} \eqref{eq: sum of transitions 3} and Lemma \ref{lem: a priori bounds 2}, we can estimate the above expressions and obtain
\begin{align}
\limsup_{ L\to \infty } \sno \Phi_{22;22}(t) \sno  \lesssim  \vert t\vert \, \vert \lambda \vert^3 \big(  \sqrt { \gamma(d,k_F)} k_F^{\frac{(d-3)}{2}} (\ln k_F)^{1/2}  + \gamma(d,k_F)  \big) .
\end{align}
This completes the derivation of inequality \eqref{eq: main estimate remark} and thus the proof of Theorem \ref{theorem: main theorem}.

\subsection{Proof of Proposition \ref{prop: lower bound for h_0} \label{sec: proof of prop}}

Set $\lambda^2 = k_F^{(2-d)}$. We first derive a lower bound for the norm difference $\sno (e^{-i h_n t } - e^{-i  \widetilde h_n  t } ) \xi_0 \sno_{\mathscr H_n}$. To this end we compute
\begin{align}\label{eq: lower bound proof line 1}
( 1 - e^{i \widetilde h_n   t } e^{-i h_n  t  } ) \xi_0  & = i \int_0^t \D s\, e^{i \widetilde h_n  s} (h_n -\widetilde h_n ) e^{- i h_n s} \xi_0 \nonumber \\
& \hspace{-1.5cm} = i t (h_n - \widetilde h_n) \xi_0 -  \int_0^t\D s \int_0^s \D r\, e^{i\widetilde h_n r }\big( \widetilde h_n   (h_n - \widetilde h_n )   -  (h_n -\widetilde h_n )   h_n \big) e^{-i h r} \xi_0.
\end{align}
With the Cauchy--Schwarz inequality we can use \eqref{eq: lower bound proof line 1} to estimate
\begin{align}\label{eq: estimate in proof of proposition:0}
& \sno (e^{-i h_n t } - e^{-i \widetilde h_n  t }) \xi_0 \sno_{\mathscr H_n}   \ge  \big\vert \lsp  \xi _0 , ( 1 - e^{i \widetilde h_n  t } e^{-i h_n t  } ) \xi_0  \rsp \big\vert \nonumber \\[2mm]
& \quad \ge  t \, \Big\vert  \sum_{i<j}^n  \lsp  \xi_0 ,  \lambda ^2 W_{k_F}(\vert y_i-y_j\vert ) \xi_0 \rsp_{\mathscr H_n} \Big\vert  \nonumber \\
& \quad \quad -  \sum_{i<j}^n \Big\vert \int_0^t \D s \int_0^s \D r \lsp e^{- i \widetilde h_n r} \xi_0, \big(\widetilde h_n \lambda ^2 W_{k_F}(\vert y_i-y_j\vert ) -  \lambda  ^2 W_{k_F}(\vert y_i-y_j\vert ) h_n \big) e^{-i h r} \xi_0 \rsp_{\mathscr H_n} \Big\vert .
\end{align} 
By condition \eqref{eq: initial states concentrated at y_1 equal y_2} we know that the first summand is bounded from below by $c_0 t $. The absolute value in \eqref{eq: estimate in proof of proposition:0}, in turn, is estimated from above by
\begin{align}\label{eq: estimate in proof of proposition}
\Big\vert \int_0^t \D s \int_0^s \D r  \dots \Big\vert \le \frac{t^2}{2} \lambda^2 W_{k_F}(0) \big( \sno \widetilde  h_n  \xi_0\sno _{\mathscr H_n} + \sno h_n \xi_0 \sno_{\mathscr H_n} \big)
\end{align}
where we used $\vert W_{k_F}(r) \vert \le W_{k_F}(0)$ and $\sno \widetilde h_n e^{- i \widetilde h_n r} \xi_0\sno= \sno \widetilde h_n  \xi_0 \sno$ and the same for $\widetilde h_n$ replaced by $h_n$. With $\lambda^2 W_{k_F}(0)\le C$ and the second assumption in \eqref{eq: initial states concentrated at y_1 equal y_2}, this implies
\begin{align}
\Big\vert \int_0^t \D s \int_0^s \D r  \dots \Big\vert  \le C t^2
\end{align}
for some constant $C>0$ independent of $k_F$ and $L$. Hence there is a time $t_1>0$ such that for all $t\in (0,t_1)$,
\begin{align} 
 \sno (e^{-ih t } - e^{-i \widetilde h_n  t }) \xi_0 \sno_{\mathscr H_n}    \ge c _0 t - \frac{ n^2}{2} C t^2 \ge   \frac{3 c _0 }{4}t .
\end{align}

In combination with Theorem \ref{theorem: main theorem} this leads to
\begin{align}
\liminf_{L\to \infty} \sno  \Psi(t) - \exp(-i \widetilde h_n t) \xi_0 \otimes e^{-iE t} \Omega_0 \sno  & \ge   \frac{3 c_0}{4}t  - C \frac{ (1+t) (\ln k_F)^{3/2}}{\sqrt {k_F}}
\end{align}
for all $t\in (0,t_1)$. Thus we can find a time $t_0$ such that for $t\in (t_0,t_1)$ and $k_F$ large enough the right side is bounded from below by $c_0 t/2$. This concludes the proof of the proposition.

\subsection{Proof of Lemma \ref{lem: lower bound effective potential}\label{sec: proof of lemma lower bound effective potential}}

For $\mu \ge 1$ we set
\begin{align}
W_{k_F}^{\le \mu}(r) & = V_d^2 \int\limits_{\vert k \vert \le k_F } \D ^d k \int\limits_{\vert l \vert \ge k_F } \D ^d l \frac{ \vert \widehat v_\infty(l-k) \vert^2 }{l^2-k^2 + (l-k)^2 + 1} \cos((l -k )\cdot r \hat a  )  \chi_{(0,\mu)} (\vert l-k\vert )   
\end{align}
where $\chi_{(0,\mu)} (\vert l-k\vert )  = 1$ for $\vert l - k \vert \le \mu$ and zero otherwise. Below we show that there is a constant $c_1>0$ such that
\begin{align}\label{eq: lower bound for W mu}
W_{k_F}^{\le \mu}( 0 )  \ge c_1 k_F^{(d-2)} .
\end{align}
Moreover, by Lemma \ref{lem: a priori bounds}, it follows that there is a constant $C_2>0$ such that
\begin{align}
\Big \vert \frac{d}{dr} W_{k_F}^{\le \mu}(r)  \Big\vert   \le \mu V_d^2 \int\limits_{\vert k \vert \le k_F } \D ^d k \int\limits_{\vert l \vert \ge k_F } \D ^d l \frac{ \vert \widehat v_\infty(l-k) \vert^2 }{l^2-k^2 + 1} \chi_{(0,\mu)} (\vert l-k\vert )   \le C_2 \mu k_F^{(d-2)}.
\end{align}
Combining the two estimates, one concludes that $ k_F^{(2-d)} W_{k_F}^{\le \mu }(r)$ is uniformly bounded from below for some small ball around $r=0$. More precisely,
\begin{align}\label{eq: lower bound for W mu 2}
\inf_{k_F\ge 1}\inf  \big\{ k_F^{(2-d)} W_{k_F}^{\le \mu }(r) \, : \, 0 \le r  \le c_1 /(2 C_2 \mu ) \big\} \ge \frac{c_1}{2}.
\end{align}

To get a lower bound for $W_{k_F}(r)$, we write
\begin{align}
W_{k_F}(r) = W_{k_F}^{\le \mu}(r) + W_{k_F}^{\ge \mu}(r)  
\end{align}
%with 
%\begin{align}
%W_{k_F}^{\ge \mu}(r)   = V_d  \int\limits_{\vert k \vert \le k_F } \D ^d k \int\limits_{\vert l \vert \ge k_F } \D ^d l \frac{ \vert \widehat v(l-k) \vert^2 }{l^2-k^2 + (l-k)^2 + 1}\chi_{(\mu,\infty)} (\vert l-k\vert )
%\end{align}
and use that $W_{k_F}^{\le \mu}(r)$ exceeds the absolute value of $W_{k_F}^{\ge \mu}(r)$ on a small ball if we choose $\mu$ large enough (since the size of this ball may shrink with $\mu$ we keep $\mu$ fixed with respect to $k_F$). To this end we shall show that there is a constant $C_3>0$ such that 
\begin{align}\label{eq: upper bound for W mu}
\sup_{k_F\ge 1} \sup_{r\ge 0} \vert k_F^{(2-d)} W_{k_F}^{\ge \mu}(r)   \vert \le C_3 \mu^{-1}
\end{align}
for all $\mu \ge 1$. Together with \eqref{eq: lower bound for W mu 2} this proves the lower bound in \eqref{eq: lower bound effective potential} since (setting $c= c_1/(2C_2 \mu)$)
\begin{align}
\inf_{k_F \ge 1}\inf_{r\in [0, c  ]}k_F^{(2-d)} W_{k_F}(r) \ge \inf_{k_F \ge 1}\inf_{r\in [0, c  ]}k_F^{(2-d)} \big( W_{k_F}^{\le \mu}(r) - \vert W_{k_F}^{\ge \mu}(r)  \vert \big) \ge \frac{c_1}{2} - C_3 \mu^{-1},
\end{align}
which is strictly positive for $\mu \ge  4 C_3/ c_1 $. 

It remains to show \eqref{eq: lower bound for W mu} and \eqref{eq: upper bound for W mu}.\medskip

\noindent \textbf{Proof of \eqref{eq: upper bound for W mu}}. Since $\vert W^{\ge \mu}_{k_F}(r) \vert \le W^{\ge \mu}_{k_F}(0)$ and $l^2-k^2 \ge k_F ( \vert l \vert - \vert k \vert )$, it is sufficient to show
\begin{align}
J =  \int\limits_{\vert k \vert \le k_F } \D ^d k \int\limits_{\vert l \vert \ge k_F } \D ^d l \frac{ \vert \widehat v_\infty(l-k) \vert^2 }{\vert l \vert - \vert k \vert + k_F^{-1} } \chi_{(\mu,\infty)} (\vert l-k\vert )\lesssim \mu^{-1} k_F^{(d-1)}.
\end{align}
This is done in Appendix \hyperlink{Appendix}{A}: Comparing with \eqref{eq: discretize J2} we see that $J = \sum_{m=1}^{M+1} J_{2,m}$ which is shown to be bounded by $J \lesssim \mu^{(d-4)} k_F^{(d-1)} + k_F^{(d-2)}$. For $k_F$ large enough this implies \eqref{eq: upper bound for W mu}.\medskip

\noindent \textbf{Proof of \eqref{eq: lower bound for W mu}}. Here we estimate 
\begin{align}
W_{k_F}^{\le \mu }(0) \ge \frac{V_d^2}{2k_F (1+R)^2} \int\limits_{\vert k \vert \le k_F} \D^d k \int\limits_{\vert l \vert \ge k_F} \D^d l \frac{1}{ \vert l \vert - \vert k \vert + k_F^{-1}} \chi_{(0,1)}(\vert k -l \vert) ,
\end{align}
where we used $\chi_{(0,\mu)}(\vert k -l \vert) \ge \chi_{(0,1)}(\vert k -l \vert)$, $\vert \widehat v(k) \vert^2 \ge 1/(1+ R)^2$ for $k^2 \le 1$, $ \vert l \vert + \vert k \vert \le 2 k_F$ and $(l-k)^2\le 1$. Next we use that the right side is bounded from below by a positive constant times
\begin{align}
 k_F^{(d-2)}  \int\limits_{k_F-1/2}^{k_F} \D s \int\limits_{ k_F   }^{s + 1/2} \D r \frac{1}{r-s+ k_F^{-1}}.\label{eq: angle integration lower bound}
\end{align}
Evaluating the remaining expression we get the desired bound, $W^{\le \mu}_{k_F}(0) \ge c_1 k_F^{(d-2)}$ for some constant $c_1 >0$ and all $\mu \ge 1$.

\section*{Appendix \hypertarget{Appendix}{A}}

\begin{lemma}\label{lem: a priori bounds 2}
Let $d\in \{1,2,3\}$, assume  \textnormal{(}\hyperlink{(A$v_L$)}{\textnormal{A}$v_L$}\textnormal{)} and let $\gamma (d,k_F)$ be defined by \eqref{def: Gamma(d)}. There exists a constant $C>0 $ \textnormal{(}depending on $R$\textnormal{)} such that for all $k_F\ge 2$,
\allowdisplaybreaks
\begin{subequations}
\begin{align}
\label{eq: lem 2 bound 5}  \lim_{ L\to\infty }    L^{-2d}\sum_{(n,k) \in T_F} \bigg( L^{-d} \sum_{l \in B_F^c }  \frac{ \vert \widehat v_L (l - k ) \vert\, \vert \widehat v_L (n-l)\vert  }{ l^2-k^2 + 1 } \bigg)^2  & \le C \gamma(d,k_F) ,   \\[1mm]
\label{eq: lem 2 bound 6} \lim_{ L\to\infty }   L^{-2d}\sum_{(l,m) \in T_F} \bigg( L^{-d} \sum_{k \in B_F}  \frac{ \vert \widehat v_L(l - k) \vert\, \vert \widehat v_L(m-k)\vert  }{ l^2-k^2 + 1 } \bigg)^2  & \le C \gamma(d,k_F) ,    \\[1mm]
\label{eq: lem 2 bound 7} \lim_{L\to \infty}  L^{-3d}\sum_{ (l,k) \in T_F } \sum_{n \in B_F}   \frac{ \vert \widehat v_L(l-k) \vert \, \vert \widehat v_L(n-k)  \vert \, \vert \widehat v_L(l-n) \vert }{ ( l^2-k^2 + 1 ) ( n^2-k^2  + 1 ) } & \le C \gamma(d,k_F) ,    \\[1mm]
\label{eq: lem 2 bound 8}  \lim_{ L\to\infty }   L^{-2d}\sum_{ k,m \in B_F} \bigg( L^{-2d} \sum_{ l, n \in B_F^c } \frac{ \vert \widehat v_L(l-k) \vert \, \vert \widehat v_L(n-m)  \vert \, \vert \widehat v_L(l-n) \vert }{ ( l^2-k^2 + 1 ) ( n^2-m^2 + 1 ) } \bigg)^2   & \le C \gamma(d,k_F)^2 ,   \\[1mm]
\label{eq: lem 2 bound 9}  \lim_{ L\to\infty } L^{-2d}\sum_{ r, l \in B_F^c} \bigg( L^{-2d} \sum_{(n,m) \in T_F} \frac{ \vert \widehat v_L(l-m) \vert \, \vert \widehat v_L (n-m)  \vert \, \vert \widehat v_L (r-n) \vert }{ ( l^2-m^2 + 1 ) ( n^2-m^2 + 1 ) } \bigg)^2 & \le C \gamma(d,k_F)^2 .   
\end{align}
\end{subequations}
\end{lemma}

In the remainder of this appendix we provide the proof of Lemmas \ref{lem: a priori bounds} and \ref{lem: a priori bounds 2}. To derive the stated bounds, we replace the expression to be estimated by its Riemann integral and proceed by estimating the latter.\medskip

\noindent \textbf{Proof of \eqref{eq: sum of transitions 1}}. The integral
\begin{align}
  J_1 = \int\limits_{\vert k \vert \le k_F} \D^d k \int\limits_{\vert l \vert \ge k_F} \D^d l \frac{1}{(1+\vert l - k \vert^2)^{2}}  
\end{align}
is bounded from above by
\begin{align}
& \int\limits_{\vert k \vert \le k_F} \D^d k \int\limits_{\vert l + k  \vert \ge k_F} \D^d l  \frac{1}{(1+\vert l  \vert^2)^{2}}  \lesssim  \int\limits_0^{k_F} \D s \, s^{d-1} \int\limits_{k_F -  s }^\infty \D r ( 1+  r)^{-2} \lesssim  k_F^{(d-1)}.
\end{align} 
%In the remainder we will frequently make use of the fact that
%\begin{align}
%\int_{\vert k \vert \le k_F } \D^d k \int_{\vert  l \vert \ge k_F } \D^d l \, \chi_{[a,b]}(\vert l - k \vert ) \le C_d k_F^{(d-1) } (b^d - a^d) (b -a)  \le C_d %k_F^{(d-1) } b^{d+1} 
%\end{align}

\noindent \textbf{Proof of \eqref{eq: sum of transitions 2}}. Let us first note estimate (for $m=1,2$)
\begin{align}
\int\limits_{\vert k \vert \le k_F} \D^d k \int\limits_{\vert l \vert \ge k_F} \D^d k \frac{ \chi_{(0,a)}(\vert k -l \vert) }{ ( \vert l \vert - \vert k \vert + k_F^{-1})^m}  \lesssim (  k_F\, a )^{(d-1)}  \int\limits_{-a}^0 \D s \int\limits_0^a \D r \frac{1}{(r-s+ k_F^{-1})^m} . \label{eq: angle integration}
\end{align}
(One of the angle integrations gives a factor $(a/k_F)^{d-1}$ while the radial component of each variable is bounded by a constant times $k_F$.) 

In \eqref{eq: sum of transitions 2} we use $l^2 - k^2 \ge k_F (\vert l \vert - \vert k \vert )$ and consider the integral
\begin{align}
  J_2   = \int\limits_{\vert k \vert \le k_F} \D^d k \int\limits_{\vert l \vert \ge k_F} \D^d l \frac{\vert \widehat v_\infty(k-l) \vert^2 }{ \vert l \vert - \vert k \vert + k_F^{-1} }.
\end{align}

To estimate such expressions, we discretize them, that is, for $M$ the smallest integer larger than $\ln k_F$, we write 
\begin{align}\label{eq: discretize J2}
J_2 =   \sum_{m = 0 }^{M+1} J_{2,m}, \quad J_{2,m} = \int\limits_{\vert k \vert \le k_F } \D^d k \int\limits_{\vert l \vert \ge k_F } \D ^dl \, \frac{\vert \widehat  v_\infty (l-k) \vert^2 }{ \vert l\vert -\vert k\vert +k_F^{-1} }  \chi_{[a_m , a_{m+1}) }(\vert l - k \vert ) ,
\end{align}
with $a_0= 0$, $a_m= \mu k_F^{( m-1 )/ M }$ for $m=1,...,M+1$ and some large enough $\mu$ (for instance $\mu \ge 10$) and $a_{M+2}=\infty$. Note that in the following estimates the value of the constant $C$ does not depend on $\mu$ or $M$.

For $m=0$ we use $\vert \widehat v_\infty(l-k) \vert \le R^{-1}$ and then apply \eqref{eq: angle integration} to get
\begin{align}
 J_{2,0}  &  \lesssim ( k_F\, \mu ) ^{(d-1)} \int\limits_{- \mu }^{0} \D s  \int\limits_{0 }^{ \mu } \D r \frac{1}{ r - s + k_F^{-1}} .
 \end{align}
Since for $a > 1 >  2\varepsilon >0$,
\begin{align}
\int\limits_{-a}^0 \D s \int\limits_{0}^{a} \D r \frac{1}{r-s + \varepsilon} 
% (2a + \varepsilon) \ln (2 a+ \varepsilon) + (b+\varepsilon) \ln (b + \varepsilon)  - (b-\varepsilon) \ln ( b - \varepsilon) - \varepsilon \ln \varepsilon %\nonumber\\[2mm]
& \le 5 a \ln(3a ) +  \varepsilon \ln (\varepsilon^{-1}),
\label{eq: simple integral 01}
\end{align}
we obtain
\begin{align}
J_{2,0} &    \lesssim  \mu^{d}(1+ \ln \mu )\, k_F^{(d-1)} .
\end{align}

For $m = 1,...,M $ we use $\vert \widehat v_\infty(l-k) \vert^2 \le \mu^{-4} k_F^{- 4 (m-1)/M}$ and then proceed with \eqref{eq: angle integration} and \eqref{eq: simple integral 01}. This gives
\begin{align}
J_{2,m}  & \lesssim   \mu^{-4} k_F^{- 4 (m-1)/M}  (k_F\,  \mu k_F^{ m/M} )^{(d-1)}  \int\limits_{- \mu k_F^{ m/M}   }^{0 } \D s  \int\limits_{0}^{ \mu k_F^{ m/M} } \D r \frac{1}{ r - s + k_F^{-1}} \nonumber \\[0mm]
& \lesssim \mu^{(d-4)} k_F^{(d-1)} k_F^{(d-4)m/M} \ln (\mu k_F^{m/M}).
\end{align}
Using $  \ln (\mu k_F^{m/M})  \lesssim  k_F^{m/(2M)}$ we can easily bound the sum of the $m$ dependent factors by
\begin{align}
\sum_{m=1}^{ M  } \big( k_F^{ (d - 7/2 )/ M } \big)^m = \frac{ k_F^{  (d-7/2)/ M }- k_F^{ (d-7/2)(M+1)/M} }{ 1 - k_F^{ (d-7/2)/ M }} \lesssim 1
\end{align}
which leads to $\sum_{m=1}^{ M  }J_{2,m} \lesssim  \mu^{(d-4)} k_F^{(d-1)}$.

%For $ J_{2,M}  $ we use $\vert \widehat v(l - k) \vert^2 \ge \mu^{-4} k_F^{-4}$ to estimate 
%\begin{align}
%J_{2,M} & \lesssim \mu^{-4} k_F^{-4  } \int_0^{k_F} \D s \, s^{d-1} \int_{k_F}^{10 \mu k_F } \D r \, r^{d-1} \frac{1}{r-s+k_F^{-1}} \notag \\[0.5mm]
%& \lesssim  \mu^{ (d-5) } k_F^{( 2d- 6 )} \bigg[ \int_{-k_F}^{0} \D s \int_{0}^{k_F} \D r \, \frac{1}{r-s+k_F^{-1}} + \int_{0}^{k_F} \D s \int_{2k_F}^{10 \mu  k_F } \D r \, \frac{1}{r-k_F+k_F^{-1}} \bigg] \notag \\[3.5mm]
%& \lesssim \mu^{(d-4)} k_F^{( 2d- 5 )  } \big(  1 +  \ln k_F \big)
%\end{align}
%which is again bounded by a constant times  $\mu^{-1} k_F^{( d- 1 ) }$ (since $d\le 3$).

In the last term we use that for $\mu$ large enough (for instance $\mu \ge 10$), we have $\vert l - k \vert \ge \vert l \vert - \vert k \vert \ge \vert l\vert /2$ and $\vert l \vert \ge 5 k_F$. By this one verifies that
\begin{align}
J_{2,M+1 }  \lesssim \int\limits_0^{k_F} \D s\,  s^{d-1}  \int\limits_{5 k_F}^\infty  \D r \,  r^{d-6}  \lesssim   k_F^{(2 d - 5 )} \lesssim k_F^{(d-2)}.
\end{align}

Combining all estimates proves \eqref{eq: sum of transitions 2}.\medskip

\noindent \textbf{Proof of \eqref{eq: sum of transitions 3}}. This bound is derived in close analogy to the previous one. The only difference is that one needs to use
\begin{align}
\int\limits_{-a}^0 \D s \int\limits_{0}^{a} \D r \frac{1}{( r-s + \varepsilon)^2 } 
% & =  2  \ln  ( a  + \varepsilon )  - \ln  (2 a  + \varepsilon ) + \ln (\varepsilon^{-1})
 \le 2 \ln (2a) + \ln (\varepsilon^{-1})  \label{eq: simple integral 02}
\end{align}
instead of \eqref{eq: simple integral 01}. We omit the details.\medskip

\noindent \textbf{Proof of \eqref{eq: sum of transitions 4}}. Since the contribution with $\vert l - k \vert\le 1 $ is smaller than the expression on the left side of \eqref{eq: sum of transitions 3}, it is sufficient to consider 
\begin{align}
J_3  =  \int\limits_{\vert k \vert \le k_F } \D^d k \int\limits_{\vert l \vert > k_F } \D ^dl \, \frac{\vert \widehat v_\infty (l-k) \vert ^2 (  l - k )^2 }{ (  l^2 -k^2 + (l-k)^2 + 1)^2 }  \chi_{[1,\infty) }(\vert l - k\vert).
\end{align}
(Note that we do not estimate the denominator as before.) We discretize again by writing
\begin{align}
J_3  = \sum_{m=1}^{M+1} J_{3,m} , \ J_{3,m}  =  \int\limits_{\vert k \vert \le k_F } \D^d k \int\limits_{\vert l \vert > k_F } \D ^dl \, \frac{\vert \widehat v_\infty (l-k) \vert ^2 (  l - k  ) ^2 }{ (  l^2 -k^2 + (l-k)^2 + 1)^2 } \chi_{[ a _m, a_{m+1} )}(\vert l - k \vert),
\end{align}
with $a_m = \mu k_F^{(m-1) / M }$ for $m=1,...,M+1$ and some large enough $\mu$, $a_{M+2} = \infty$ and $M$ the smallest integer larger than $\ln k_F$.

For $m = 1,...,M$ we use $l^2-k^2+(l-k)^2+1 \ge k_F (\vert l \vert - \vert k \vert + k_F^{-1})$ and $\vert \widehat v_\infty (l-k) \vert ^2 \vert  l - k \vert^2  \lesssim ( \mu k_F^{(m-1)/M})^{-2}$. By means of \eqref{eq: angle integration} and \eqref{eq: simple integral 02} we then get
\begin{align}
J_{3,m} & \lesssim  k_F^{(d-3)} (\mu k_F^{(m-1)/M})^{-2} (\mu k_F^{m/M})^{(d-1)}    \int\limits_{ -\mu k_F^{m/M}  }^0 \D s   \int\limits_0^{ \mu k_F^{m/M} } \D r \frac{1}{(r-s+k_F^{-1})^2} \notag \\[0mm]
&  \lesssim \mu^{(d-3)} k_F^{(d-3)} k_F^{(d-3)m/M} \ln ( \mu k_F ).
\end{align}
The sum over the $m$ dependent factors is bounded by
\begin{align}
\sum_{m=1}^{M}  k_F^{ m(d-3) /M } \le M \le 1+ \ln k_F
\end{align}
and thus we have $\sum_{m=1}^{M} J_{3,m}  \lesssim  \mu^{(d-3)} k_F^{(d-3)}(1+\ln k_F)^2$. 

In the last term we bound the factor $(l-k)^2$ by the denominator. This gives
\begin{align}
J_{3,M+1}  & \lesssim \int\limits_{\vert k \vert \le k_F } \D^d k \int\limits_{\vert l \vert > k_F } \D ^dl \, \frac{\vert \widehat v_\infty (l-k) \vert ^2 }{   l^2 -k^2 + (l-k)^2 + 1  } \chi_{[ \mu k_F , \infty )}(\vert l - k \vert) .
\end{align}
From here, we use $l^2-k^2\ge k_F^{-1}(\vert l \vert - \vert k \vert)$ and $\vert l - k \vert \ge \vert l \vert - \vert k \vert \ge \vert l\vert /2$ and $\vert l \vert \ge 5 k_F$ (which is true for instance for $\mu \ge 10$). Hence we get
\begin{align}
J_{3,M+1} & \lesssim  k_F^{-1} \int\limits_{\vert k \vert \le k_F } \D^d k \int\limits_{\vert l \vert > k_F } \D ^dl \, \frac{\vert \widehat v_\infty (l-k) \vert ^2 }{   \vert l \vert - \vert k \vert +k_F^{-1} } \chi_{[ \mu k_F , \infty )}(\vert l - k \vert) \notag \\ 
& \lesssim  k_F^{-1} \int_0^{k_F} \D s\, s^{d-1}  \int_{5 k_F}^\infty  \D r \,  r^{d-4}   \lesssim  k_F^{(2 d - 4 )} .
\end{align}
 
\noindent \textbf{Proof of \eqref{eq: lem 2 bound 5}}. To estimate the integral
\begin{align}
J_4 & =  \int\limits_{\vert k \vert \le k_F } \D ^d k \int\limits_{\vert n \vert \ge k_F } \D ^d n  \bigg(\,  \int\limits_{\vert l \vert \ge k_F} \D^d l \frac{ \vert \widehat v_\infty (l - n) \vert\, \vert \widehat v_\infty (l-k)\vert  }{ l^2-k^2 + 1 }\, \bigg) ^2  ,
\end{align}
we employ
\begin{align}
\int_{\mathbb R^d} \D^d n \vert \widehat v_\infty (l-n) \vert \ \vert \widehat v_\infty (l'- n )  \vert \lesssim 1.
\end{align}
%To verify the latter, we use for $ \vert l - l ' \vert \ge 1$ the identity
%\begin{align*}
%\int_{\mathbb R^d}\D^d n \frac{1}{\vert n - l \vert^2} \frac{1}{\vert n - l'\vert ^2 } = \frac{ C_d }{\vert l - l' \vert ^{4-d}},
%\end{align*}
%whereas for $\vert l - l' \vert \le 1$, we compute
%\begin{align}
%\int_{\mathbb R^d}  \D^d n \vert \widehat v_\infty ( n - l ) \vert \, \vert \widehat v_\infty (n-l') \vert &    = 
%\int_{\mathbb R^d} \D^d n \frac{1}{1+\vert n \vert^2} \, \frac{1}{1+ \vert  n - ( l' - l)  \vert ^2 }   \notag  \\[1mm]
%&  \lesssim   \int_{\mathbb R^d} \D^d n \, \bigg[ \frac{\chi_{[0,2]}(\vert n \vert ) }{1+\vert n \vert^2} \,  +\,  \frac{\chi_{(2,\infty)}(\vert n \vert) }{\vert n \vert^4}  \bigg]  \lesssim   1.
%\end{align}
Thus we have
\begin{align} \label{eq: integral estimate}
J_4 \lesssim \int\limits_{\vert k \vert \le k_F } \D ^d k  \bigg(   \int\limits_{\vert l \vert \ge k_F} \D^d l \frac{ \vert \widehat v_\infty (l-k)\vert  }{ l^2-k^2 + 1 } \bigg) ^2 \lesssim k_F^{-2} \int\limits_{\vert k \vert \le k_F} \D ^d k  \bigg(  \int\limits_{\vert l \vert \ge k_F}   \D^d l \frac{ \vert \widehat v_\infty (l-k) \vert   }{ \vert l \vert - \vert k \vert  + k_F^{-1} }  \bigg) ^2  .
\end{align}
In the last expression we estimate the integrand by
\begin{align}
 \bigg(  \int\limits_{\vert l \vert \ge k_F}   \D^d l \frac{ \vert \widehat v_\infty (l-k) \vert   }{ \vert l \vert - \vert k \vert  + k_F^{-1} }  \bigg) ^2  & \lesssim    A_0(k)^2 + \bigg(\sum_{m=1}^{M} A_m(k) \bigg)^2  + A_{M+1}(k)^2  
\end{align}
with 
\begin{align}
A_m(k) = \int\limits_{\vert l \vert \ge k_F}  \D^d l \frac{ \vert \widehat v_\infty (l-k)  \vert   }{ \vert l \vert - \vert k \vert  + k_F^{-1} } \chi_{ [ a_m , a_{m+1} ) }(\vert l - k \vert) 
\end{align}
where $M$ is the smallest integer larger than $\ln k_F$, and 
\begin{align}\label{eq: a_m}
a_0 = 0, \quad a_m = 1 0 k_F^{(m-1) /M} \quad (m=1,...,M+1), \quad a_{M+2}=\infty .
\end{align}

The $A_m(k)$ are estimated by
\begin{align}
A_0(k) & \lesssim   \int\limits_{k_F}^{k_F+10} \D r \frac{1}{r- \vert k \vert +k_F^{-1}} \chi_{[-10,0)}(\vert k \vert - k_F) \notag \\
& \lesssim \big( \ln (k_F+ 10  - \vert k \vert  + k_F^{-1} ) - \ln (k_F - \vert k \vert  + k_F^{-1} )\big) \chi_{[- 10 ,0)}(\vert k \vert - k_F) \notag \\[4mm]
& \lesssim \big( \ln (2 0+  k_F^{-1} ) - \ln ( k_F^{-1} )\big) \chi_{[-10,0)}(\vert k \vert - k_F)  \lesssim \ln k_F\,  \chi_{[-10,0]}(\vert k \vert - k_F),
\end{align}
which leads to
\begin{align}
k_F^{-2} \int\limits_{\vert k \vert \le k_F} \D^d k A_0(k)^2 \lesssim k_F^{(d-3)} (\ln k_F)^2.
\end{align}

For $m = 1,...,M $ we use $\vert \widehat v_\infty (l-k) \vert \lesssim k_F^{-2m/M}$ and get
\begin{align}
A_m(k) & \lesssim  k_F^{(d-1)m/M}  k_F^{-2m/M} \int\limits_{k_F}^{k_F+10 k_F^{ m/M} } \frac{1}{r- \vert k \vert +k_F^{-1}} \chi_{[-10 k_F^{ m/M},0) }(\vert k \vert - k_F) \notag \\[0mm]
& \lesssim  k_F^{(d-3) m/M } \big( \ln (2 0 k_F^{ m/M} +  k_F^{-1} ) - \ln ( k_F^{-1} )\big) \chi_{[- 10 k_F^{ m/M} ,0 ) }(\vert k \vert - k_F) \notag \\[2.5mm]
& \lesssim k_F^{(d-3) m/M }  \ln k_F \, \chi_{[- 10 k_F^{ m/M} ,0) }(\vert k \vert - k_F).
\end{align}
We thus find
\begin{align}
k_F^{-2 }\int\limits_{\vert k \vert \le k_F} \D^d k \bigg(\sum_{m=1}^{M} A_m(k) \bigg)^2  & \lesssim k_F^{(d-3)}  (\ln k_F)^2 \sum_{  n ,  m = 1 }^{M }k_F^{(d-3) m/M } k_F^{(d-2) n/M } \notag \\
& \lesssim  \begin{cases}  k_F^{(d-3)} (\ln k_F)^3 \quad (d=1,2)\\[1mm]
  k_F (\ln k_F)^3  \quad  \ \quad (d=3)\end{cases}.
\end{align}

In the last summand we use $\vert l \vert \ge 5k_F$ to get
\begin{align}
\int\limits_{\vert k \vert \ge k_F} \D^d k A_{M+1}(k)^2  \le k_F^{-2}  
\int\limits_{\vert k \vert \ge k_F} \D^d k \bigg( \int_{5k_F}^\infty r^{d-4} \D r \bigg)^2 \lesssim k_F^{3d-8}.
\end{align}

\noindent \textbf{Proof of \eqref{eq: lem 2 bound 6}}. The derivation is almost the same as the previous one. Consider
\begin{align}
J_{5} & =  \int\limits_{\vert m \vert \le k_F } \D ^d m \int\limits_{\vert l \vert \ge k_F} \D ^d l   \bigg( \int\limits_{\vert k \vert \le k_F} \D^d k \frac{ \vert \widehat v_\infty (k - m ) \vert\, \vert \widehat v_\infty (l-k)\vert  }{ l^2-k^2 + 1 } \bigg)^2 \notag \\
&  \lesssim k_F^{-2} \int\limits_{\vert l \vert \ge k_F } \D ^d l  \bigg(  \int\limits_{\vert k \vert \le k_F}  \D^d k \frac{ \vert \widehat v_\infty (l-k)\vert   }{ \vert l \vert - \vert k \vert  + k_F^{-1} }  \bigg) ^2 \notag \\
& \lesssim k_F^{-2} \int\limits_{\vert l \vert \ge k_F} \D ^d l  \bigg(   B_0(l)^2 + \bigg(\sum_{m=1}^{M} B_m(l) \bigg)^2 + B_{M+1}(l)^2  \bigg) ,
\end{align}
with
\begin{align}
B_m(l) = \int\limits_{\vert l \vert \ge k_F}  \D^d k \frac{ \vert \widehat v_\infty (l-k) \vert   }{ \vert l \vert - \vert k \vert  + k_F^{-1} } \chi_{[a_m,a_{m+1})}(\vert l - k \vert) 
\end{align}
and $(a_m)_{m=0}^{M+2}$ defined by \eqref{eq: a_m}. 

The $B_m(l)$ are estimated by
\begin{align}
B_0(l) & \lesssim \int\limits_{k_F-10}^{k_F}\D s \frac{1}{ \vert l \vert  -s +k_F^{-1}} \chi_{[0,10)}(\vert l \vert - k_F) \notag \\
& = \big( \ln ( \vert l \vert - k_F + 10  + k_F^{-1} ) - \ln (\vert l \vert - k_F + k_F^{-1} )\big) \chi_{[0, 10 )}(\vert l \vert - k_F) \notag \\[3.5mm]
& \lesssim \big( \ln (2 0 +  k_F^{-1} ) - \ln ( k_F^{-1} )\big) \chi_{[-a_1,0)}(\vert k \vert - k_F)  \lesssim  \chi_{[-10,0)}(\vert k \vert - k_F) \ln k_F ,
\end{align}
which leads to
\begin{align}
 k_F^{-2} \int\limits_{\vert l \vert \ge k_F} \D ^d l \,   B_0(l)^2   \lesssim k_F^{(d-3)} (\ln k_F)^2.
\end{align}

For $m = 1,...,M$, we have
\begin{align}
B_m(l) & \lesssim  k_F^{(d-3)m/M} \int\limits_{k_F - 10k_F^{m/M}}^{k_F} \D s \frac{1}{\vert l \vert - s  +k_F^{-1}} \chi_{[0,   10k_F^{m/M} ) }(\vert l \vert - k_F) \notag \\[2mm]
& \lesssim k_F^{(d-3)m/M} \chi_{[0,  10k_F^{m/M}  ) }(\vert l \vert - k_F)\ln k_F  
\end{align}
and thus get
\begin{align}
k_F^{-2 }\int\limits_{\vert l \vert \ge  k_F} \D^d l \bigg(\sum_{m=1}^{M } B_m(l) \bigg)^2  &  \lesssim k_F^{(d-3)}  (\ln k_F)^2 \sum_{n,m =1}^{M}  k_F^{(d-3)m/M} k_F^{(d-2)m/M}  \notag\\
& \lesssim \begin{cases} k_F^{(-2)} (\ln k_F)^3 \quad\quad\ \,  (d=1)\\[2mm]
 k_F^{2d-5} (\ln k_F)^2  \quad  \ \quad (d=2,3)\end{cases}.
\end{align}

In $B_{M+1}(l)$ we use $\vert l \vert \ge 5 k_F$ and proceed by
\begin{align}
k_F^{-2 }\int\limits_{\vert l \vert \ge  k_F} \D^d l B_{M+1}(l) \lesssim k_F^{-2} \int\limits_{\vert l \vert \ge k_F } \D ^d l \bigg( k_F^d \frac{1}{\vert l \vert^3} \bigg)^2\lesssim k_F^{(3d-8)} \lesssim \begin{cases}  k_F^{( d - 4 ) } \quad (d=1,2) \\
 k_F \quad \quad \ (d=3)\end{cases} .
\end{align}

\noindent \textbf{Proof of} \eqref{eq: lem 2 bound 7}-\eqref{eq: lem 2 bound 9}. Here we can use $\vert \widehat v_\infty(l-n) \vert \le R^{-1}$ and
\begin{align}
J_6 & = \int\limits_{\vert k \vert \le k_F} \D^d k  \int\limits_{\vert l \vert \ge k_F} \D^d l  \int\limits_{\vert n \vert \ge k_F}  \D ^d n \frac{ \vert \widehat v_\infty (l-k) \vert \, \vert \widehat v_\infty (n-k)  \vert \, \widehat v_\infty(l-n) \vert }{ ( l^2-k^2 + 1 ) ( n^2-k^2  + 1 ) } \notag \\
& \lesssim \int\limits_{\vert k \vert \le k_F} \D^d k  \Bigg( \int\limits_{\vert l \vert \ge k_F} \D^d l   \frac{ \vert \widehat v_\infty (l-k) \vert }{ l^2-k^2 + 1} \Bigg)^2
\end{align}
which has been estimated in \eqref{eq: integral estimate}. 

Similarly, also for
\begin{align}
J_7 & = \int\limits_{\vert k \vert \le k_F}  \D^d k \int\limits_{\vert m \vert \le k_F} \D ^d m \Bigg( \int\limits_{\vert  l \vert \ge k_F} \D^d l  \int\limits_{\vert n \vert \ge k_F} \D ^d n \frac{ \vert \widehat v_\infty (l-k) \vert \, \vert \widehat v_\infty (n-m)  \vert \, \widehat v_\infty (l-n) \vert }{ ( l^2-k^2 + 1 ) ( n^2-m^2 + 1 ) }  \bigg)^2 \notag \\
&   \le \Bigg( \int\limits_{\vert k \vert \le k_F}  \D^d k \Bigg( \int\limits_{\vert l \vert \ge k_F} d^d l   \frac{ \vert \widehat v_\infty (l-k) \vert \,  }{ l^2-k^2 + 1 } \Bigg)^2 \Bigg)^2 
\end{align}
and
\begin{align}
J_8 & =   \int\limits_{\vert r \vert \ge k_F} \D ^d r \int\limits_{\vert l \vert \ge k_F } \D ^d l  \Bigg(\int\limits_{\vert n \vert \ge k_F} \D ^d n  \int\limits_{\vert m \vert \le k_F} \D ^d m \frac{ \vert \widehat v_\infty(l-m) \vert \, \vert \widehat v_\infty(n-m)  \vert \, \widehat v_\infty (r-n) \vert }{ ( l^2-m^2 + 1 ) ( n^2-m^2 + 1 ) }  \Bigg)^2 \notag \\
& \le \Bigg( \int \limits_{\vert m \vert \le k_F}  \D^d m \Bigg( \int_{\vert  l \vert \ge k_F} \D^d l   \frac{ \vert \widehat v_\infty (l-m) \vert \,  }{ l^2- m ^2 + 1}  \Bigg)^2 \Bigg)^2.
\end{align}

\section*{Acknowledgments}

We are grateful to Maximilian Jeblick for his contributions at an earlier stage of this project.

\vfill 

\vspace{7mm}
\noindent (David Mitrouskas)\\
\textsc{Institute of Science and Technology (IST) Austria\\
Am Campus 1, 3400 Klosterneuburg, Austria}\\
\textit{E-mail address:} \texttt{david.mitrouskas@ist.ac.at}\\

\noindent (Peter Pickl)\\
\textsc{Fachbereich Mathematik, Universit\"at T\"ubingen\\
Auf der Morgenstelle 10, 72076 T\"ubingen, Germany}\\
\textit{E-mail address:} \texttt{p.pickl@uni-tuebingen.de}

\end{spacing}

\end{document}